\documentclass[12pt]{iopart}

\usepackage{graphicx}
\usepackage{dcolumn}
\usepackage{bm}
\usepackage{amssymb}%

\newcommand{\be}{\begin{eqnarray}}
\newcommand{\ee}{\end{eqnarray}}
\begin{document}

\title[Observational Testability of Kerr bound in X-ray Spectrum of BH Candidates]{Observational Testability of Kerr bound in X-ray Spectrum of Black-Hole Candidates}

\author{Rohta Takahashi}
\address{ 
Cosmic Radiation Laboratory, the Institute of Physical and Chemical Research, 
2-1 Hirosawa, Wako, Saitama 351-0198, Japan}
\ead{rohta@riken.jp}

\author{Tomohiro Harada}
\address{
Department of Physics, Rikkyo University, Toshima, Tokyo 171-8501, Japan}
\ead{harada@rikkyo.ac.jp}

\begin{abstract}
The specific angular momentum of a Kerr black hole must not be larger 
than its mass. The observational confirmation of this bound which 
we call a Kerr bound directly suggests the existence of a black hole. 
In order to investigate observational testability of this bound by using the 
X-ray energy spectrum of black hole candidates, we calculate 
energy spectra for a super-spinning object (or a naked singularity) which 
is described by a Kerr metric but  whose specific angular momentum is larger 
than its mass, and then compare the spectra of this object with those of a black hole. 
We assume an optically thick and geometrically thin disc around the 
super-spinning object and calculate its thermal energy spectrum seen by 
a distant observer by solving general relativistic 
radiative transfer equations including usual special and general 
relativistic effects such as Doppler boosting, gravitational 
redshift, light bending and frame-dragging. Surprisingly, for 
a given black hole, we can always find its super-spinning counterpart 
with its spin $a_*$ in the range $5/3<a_{*}<8\sqrt{6}/3$ 
whose observed spectrum is very similar to and practically 
indistinguishable from that of the black hole. As a result, we conclude 
that to confirm the Kerr bound we need more than the X-ray thermal 
spectrum of the black hole candidates. 
\end{abstract}


\maketitle

\section{Introduction}
From the last century, many black hole candidates are observationally 
discovered and it is believed that the spacetime around the object  
is well described by a Kerr metric. In the Kerr metric, when a central 
object is a black hole, it is required that its specific angular momentum 
should not be larger than its mass \cite{mtw73,fn98}, i.e. $|a|\le M$. 
Here, $a$ denotes a specific angular momentum of a black hole 
and is defined as $a=J/M$ where $M$ and $J$ are the mass and 
the angular momentum of the black hole, respectively. 
In this case, an event horizon exists around the black hole. We call this bound  
a Kerr bound \cite{gh09,bf09,bft09}. The observational confirmation of 
this bound leads to the confirmation of the existence of a black hole. 

In case that the specific angular momentum of a central object is larger than
its mass, curvature singularity where spacetime curvature
diverges is not surrounded by an event horizon. To avoid this, the 
cosmic censorship conjecture in which spacetime singularity should
be concealed from our world is proposed \cite{p69,p79,w97,p98}. 
In some numerical simulations which start from configurations similar
to a star before gravitational collapse, an apparent horizon
forms before the formation of spacetime curvature singularity and
a final object inevitably becomes a black hole \cite{ss04}. 
On the other hand, the appearance of curvature singularity 
without being surrounded by an apparent horizon has been suggested
in the numerical simulation of axisymmetric gravitational collapse of collisionless 
particles \cite{st91, st92}. It has been also revealed that 
curvature singularity not surrounded by a horizon appears in the 
spherically symmetric collapse in the critical and super-critical 
cases (\cite{harada04} and references therein). 
Of course, it is also well known that in cylindrically symmetric systems 
a black hole cannot be formed after gravitational collapse
\cite{t65,t72}, and it is not fully
excluded that the spacetime singularity which is not surrounded
by a horizon forms when the gravitational
collapse starts from nearly cylindrically symmetric systems 
\cite{b02}. However, it is questionable that such 
highly symmetric configuration forms in any astrophysical situations.  
We can also see that most examples of spacetime singularities
not surrounded by a horizon can be regarded as precursory or transient
singularities followed by the formation of event and/or apparent horizons. It
is also discussed that spacetime singularity not surrounded by a 
horizon is dynamically unstable\cite{cpcc08a,cpcc08b,pccc09}. 
Moreover, the object with its specific angular
momentum larger than its mass evolves to a black hole through 
accretion processes of mass and angular momentum from a rotating disk around 
the object \cite{mtw73,t74}. That is, it would be a physically reasonable 
assumption that the black hole candidates with mass accretion observed so far are 
really black holes (although recently, another scenario has been discussed, see \cite{js09}). 
Recently, the numerical simulations of hydrodynamic accretion onto a super-spinning 
object have suggested that for the object with 
its spin $a/M$ slightly larger than unity the mass accretion is prohibited 
because of the repulsive force near the central object due to  
the spacetime geometrical effects \cite{bfh09}. 

On the other hand, in past studies many attempts were performed to obtain
direct evidence of the existence of a black hole from observational data 
(see \S\ref{sec:dis}). Especially, thanks to the developments of radio interferometers 
achieving highest spatial resolution in existing telescopes, the black hole in the 
Galactic Center Sgr A* with a large apparent size begins to be spatially resolved
\cite{s05,d08}. The spatial resolution of these radio
interferometers is comparable to the size of the apparent size of a black
hole shadow in the nearby galactic centers such as Sgr A*
\cite{b09,y09,h09} and M87 \cite{bl09}. The data recently obtained
by sub-millimeter interferometers \cite{d08} contain information of the size of 
luminous matters whose size is comparable to the size of the event horizon of the 
black hole in Sgr A* \cite{b09,y09,h09}. Although as described above the
Kerr bound is assumed to be valid by many authors, we have not yet obtained the final 
confirmation of the bound from observational data. Even for the black hole in Sgr A* 
in which there are plenty of observational data such as energy spectrum, linear 
and circular polarization, radio visibility and light curves in the wide range of 
observed frequencies, we have not yet obtained the final value of the spin of the black hole 
in Sgr A* \cite{b09,y09,h09} and the Kerr bound is not also observationally 
confirmed. 

Recently, several theoretical studies relating to the confirmation of the Kerr 
bound are performed for a black hole which we can potentially directly image 
in the near future such as the case of Sgr A* and in these studies the apparent size 
of a central object is estimated \cite{bf09,bft09,hm09}. The strategy they adopted 
is as follows. They first assume a central object which is described by the Kerr metric 
but have specific angular momentum larger than its mass. Next, they calculate 
the observational signatures of the assumed object. Finally, they compare the observational 
signatures with those of a black hole. As observational signatures, the apparent shapes 
and sizes of the assumed objects are calculated. From these calculations, 
the very large values of the spin of the central object in Sgr A* are ruled out 
from its large size \cite{bf09,hm09}. They assume that general relativity is not valid at 
a central region near the curvature singularity but replaced by an alternative theory such as quantum gravity theory (see also \cite{gh09}).  

In order to spatially resolve the apparent image of a black hole by future interferometers, 
its apparent angular size should be larger than several micro-arcseconds, which correspond 
to the spatial resolution of radio interferometers in the near future \cite{d08}. 
Even with these highest spatial resolution, most of the black hole candidates discovered 
in X-ray observations can not be spatially resolved because of their too small angular sizes. 
For most of the black holes discovered in X-ray, energy spectrum and light curves 
are observationally obtained. For some of these black holes observed in X-ray, 
the parameters of the black hole such as the mass and the spin are determined by 
the spectral fittings by assuming that the object is a black hole 
\cite{zcc97,gmne01,dbht05,msnrdl06,smndlr06,rm06,nms08,mns08}. 
For these black hole candidates, it is open to question whether it is possible to confirm the Kerr 
bound from the observational data. The main purpose of the present study is to answer this. 
In order to do this, we take the same strategy as described in the last paragraph. That is, 
we first consider the object which violates the Kerr bound, i.e. with $a/M>1$. In this paper, 
we call this object  a super-spinar \cite{gh09,bfh09}. 
Next, the energy spectrum of the assumed object is calculated and finally compare the 
spectrum with that of a black hole. The X-ray spectrum of a black hole candidate generally 
consists of a thermal component originating from the accretion disc around the black hole 
and a non-thermal component originating from high energy photons which are up-scattered 
in e.g. corona above or in the accretion disc. In this study, we only assume a thermal 
component for simplicity. The observed energy spectrum is calculated by solving the general 
relativistic radiative transfer including usual special and general relativistic effects such as 
Doppler boosting, gravitational redshift, light bending and frame-dragging. 
In this paper, we assume no emission from a central object. 

The present paper is organized as follows. In \S\ref{sec:disc}, physical assumptions and 
disc structure are given. In \S\ref{sec:spec}, we calculate the local radiation flux, the radial 
temperature profile and the energy spectrum of the disc. We give discussion in \S\ref{sec:dis} 
and conclusions are presented in \S\ref{sec:con}. 
Throughout this paper, we use the geometrical units $c=G=1$. 

\section{Structure of an Accretion disc}
\label{sec:disc}

In this section, we describe the basic disc structure used in the calculations in the next 
section. Although some part of this section was already investigated in the past studies 
\cite{df74, df78, cn79, rt79, s80, s81a, s81b, bsb89}, 
we describe these for the completeness of the description 
and the preliminaries for the next section. In the process of the mass accretion onto the 
central compact object such as a black hole or a super-spinar, when the accreting 
matters have some angular momentum, accreting fluid usually forms disc-like structure 
around a compact object. This astrophysical object is called an accretion disc. In the 
accretion disc, the angular momentum of the fluid is transferred outward due to the 
viscous stress caused by magnetic and/or turbulent effects. Then, matters are allowed 
to gradually and spirally accrete inward. On the other hand, the viscous stress converts 
the gravitational energy of the accreting matters into other forms of energy such as the 
thermal, radiation and/or magnetic energies. In the present study, 
we consider the geometrically thin and optically thick accretion disc,
where the gravitational energy is effectively released as the radiation energy, and then 
produce the significant radiation which can be observed. According to the accretion 
disc theory, this type of disc structure is achieved for the accretion disc whose mass 
accretion rate is nearly sub-Eddington, i.e. $\dot{M}< L_{\rm Edd}/c^2$. For the 
geometrically thin disc, the disc thickness $H$ in the vertical direction at some radius 
$r$ is much smaller than the radius $r$, i.e. $H\ll r$. In terms of this type of the accretion 
disc, the general relativistic disc around the Kerr black hole is given in \cite{nt73} and 
\cite{pt74}. In the present study, by adopting the basically same calculation methods and 
assumptions in \cite{pt74}, we calculate the disc structure around the super-spinar. 

We neglect the self-gravity of the accretion disc. As a background
geometry, we consider the stationary, axisymmetric and asymptotically flat spacetime 
described as 
\begin{equation}
ds^2=-e^{2\nu}dt^2+e^{2\psi}(d\varphi-\omega dt)^2+e^{2\mu_1}dr^2+e^{2\mu_2}d\theta^2, 
\label{eq:metric}
\end{equation}
where 
\begin{eqnarray}
e^{2\nu}=\Sigma\Delta/A,~~e^{2\psi}=\sin^2\theta A/\Sigma,~~e^{2\mu_1}
=\Sigma/\Delta,~~e^{2\mu_2}=\Sigma,~~\omega=2Mar/A. \nonumber \\
\end{eqnarray}
Here, $M$ is the mass of the central object, $a$ is its angular momentum
per unit mass, $\omega$ is the angular velocity of the frame-dragging
around the central object, and the functions $\Delta$, $\Sigma$ and $A$
are defined as $\Delta=r^2-2Mr+a^2$, $\Sigma=r^2+a^2\cos^2\theta$, and
$A=(r^2+a^2)^2-a^2\Delta\sin^2\theta$, respectively. The horizon radius 
$r_{\rm H}$ of the black hole is given by $r_{\rm H}/M=1+\sqrt{1-a_*^2}$
where $a_*\equiv a/M$ for $-1\le a_* \le 1$. 
While for $-1\le a_* \le 1$ the central object is a black hole, 
for $a_* < -1$ or $1< a_*$ it is a super-spinar. In this study, we
mainly focus on the super-spinar with $1<a_{*}$. 

We assume that the matters in the disc move in equatorial circular
geodesic orbits around the central object as in \cite{pt74} and \cite{bpt72}. Such
orbits of the particles in the Kerr geometry are described by the three
constants of motion; the total energy $E$, the angular momentum parallel to
symmetry axis $L$ and the Carter constant $Q$ \cite{bpt72, c68}. The
radial component of the equations governing the orbital trajectory is
given by $\Sigma dr/d\lambda=\pm [V_r(r)]^{1/2}$, where $\lambda$ is the
parameter along the trajectory and
$V_r(r)=[E(r^2+a^2)-La]^2-\Delta[r^2+(L-aE)^2+Q]$ in \cite{bpt72}. The
energy and the angular momentum of the particle with the circular orbit
on the equatorial plane are calculated from the conditions $V_r(r)=0$ and 
$V_r^\prime(r)=0$ and are given by 
\begin{eqnarray}
E&=&s_1[r^{1/2}(r-2M)+s_2 aM^{1/2}]/p, \label{eq:E}\\ 
L&=&s_1 s_2 M^{1/2}(r^2+a^2-s_2 2M^{1/2}ar^{1/2})/p, \label{eq:L}
\end{eqnarray}
where $p=r^{3/4}(r^{3/2}-3Mr^{1/2}+s_2 2aM^{1/2})^{1/2}$, $s_1=\pm 1$
and $s_2=\pm 1$.  
In the limit of $r\to \infty$, the sign of the energy becomes 
positive (negative) for $s_1=1$ ($-1$). 
It is noted that the sign $s_1 = -1$ corresponds to negative energy as measured 
by local observers and gives unphysical solutions in the context of the present 
paper (for details, see \cite{bsb89}). 
On the other hand, 
$s_2=1$ and $-1$ respectively correspond to equatorial circular orbits of 
the 1st family and the 2nd family \cite{s80}. For the orbits with $s_1=1$,  
the 2nd family orbit always corresponds to a retrograde orbit. 
In the case of a black hole ($a_*\le 1$), the 1st family orbit always corresponds 
to a prograde orbit outside the horizon \cite{bpt72}. 
On the other hand, in the case of a super-spinar ($a_*>1$)  the 1st family orbit 
can be a retrograde orbit near the super-spinar, while far from the super-spinar 
the 1st family orbit always corresponds to a prograde orbit. These features are 
investigated in the past studies \cite{df74,s80}. 
From the energy $E(=-u_t)$ and the angular momentum $L(=u_\phi)$
given above, all components of the four-velocity $u^\mu$ of the particle
with a circular orbit in the equatorial plane can be calculated. For
this orbit, the angular velocity $\Omega(=u^\phi/u^t)$ is given by
$\Omega=s_2M^{1/2}/[r^{3/2}+s_2 aM^{1/2}]$. This is the Keplerian
angular velocity in the Kerr geometry. By using these $E$ and $L$, 
the marginally bound circular orbit $r_{\rm mb}$ and the ISCO $r_{\rm ISCO}$ are obtained 
from $E/\mu=1$ and $V_r^{\prime\prime}(r)=0$, respectively, as in \cite{bpt72}. 
The circular orbit exists for the case that the denominator of Eqs. (\ref{eq:E}) and 
(\ref{eq:L}) is real, i.e. $p\ge 0$. The limiting case, $p=0$, gives an orbit with infinite 
energy per unit rest mass, and hence 
the radius of the photon circular orbit $r_{\rm ph}$ \cite{bpt72}.  

\begin{figure}
\begin{center}
\vspace{+0mm}
\hspace*{-0mm}
\includegraphics[width=120mm]{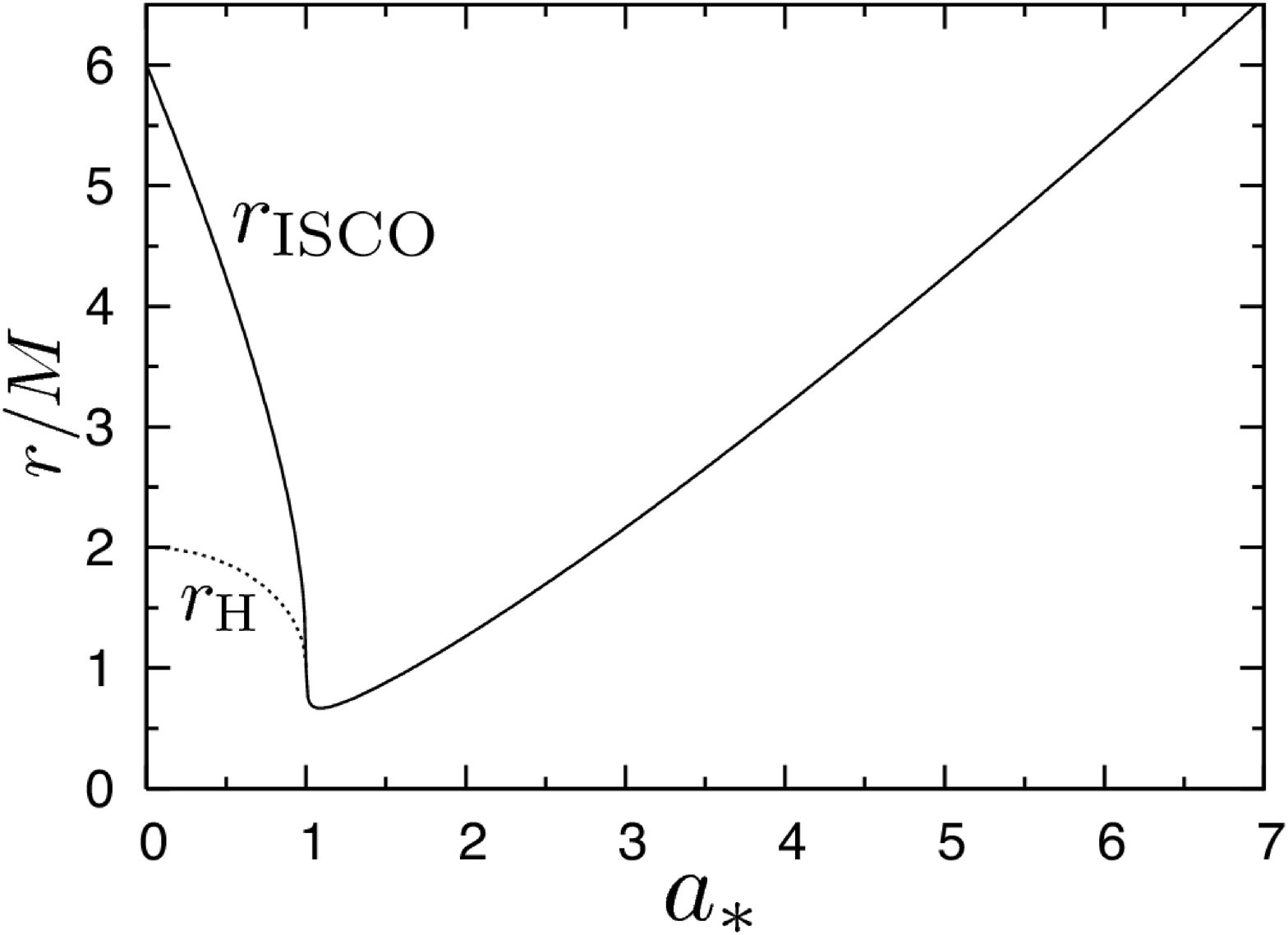}
\caption{\label{fig:superKerr_radius} 
The radii of the event horizon $r_{\rm H}$ ({\it dotted line}) and the innermost stable 
circular orbit (ISCO) $r_{\rm ISCO}$ ({\it solid line}) for a black hole ($a_*\leq 1$) and 
a super-spinar ($a_*>1$) as a function of the Kerr parameter $a_*$. Only the 
quantities corresponding to the direct motions are shown.   
}
\end{center}
\end{figure}

For the accretion disc consisting of the materials with the circular
orbit, the inner boundary with no torque is usually assumed. While
outside the inner boundary the gravitational energy is effectively released as the 
radiation, inside the inner boundary the matters fall freely onto the central object with the 
energy and the angular momentum with the values at the inner
boundary. That is, inside this radius which is called the plunging
region, there is no radiation. The recent three-dimensional
magnetohydrodynamic simulations around the rotating black holes support
this assumption \cite{s08}. In this study, we assume that the viscous torque vanishes 
at the ISCO, $r_{\rm ISCO}$, where $dE/dr=dL/dr=0$. For the rotating black hole, the 
analytic expression for $r_{\rm ISCO}$ is given by Eq. (2.21) in \cite{bpt72}. For any 
values of the spin $a_*$, the radius of ISCO is analytically given by
\begin{eqnarray}
r_{\rm ISCO}/M&=&3+Z_2-{\rm sgn}_2\left[(3-Z_1)(3+Z_1+2Z_2)\right]^{1/2},\\
Z_1&=&1+\left|1-a_*^2\right|^{1/3}\left[\left|1-|a_*|\right|^{1/3}+{\rm sgn}_1(1+|a_*|)^{1/3}\right],\nonumber\\
Z_2&=&\left(3a_*^2+Z_1^2\right)^{1/2},\nonumber\\
{\rm sgn}_1&=&\left\{
	\begin{array}{ll}
	+1 & ~{\rm for}~~a_*^2\leq 1~~({\rm black~hole})\\
	-1 & ~{\rm for}~~a_*^2> 1~~({\rm naked~singularity})\\
	\end{array}
	\right.
	 \nonumber\\
{\rm sgn}_2&=&\left\{
	\begin{array}{ll}
	+1 & ~{\rm for}~~a_*\geq 0 \\
	-1 & ~{\rm for}~~a_*<0.\\
	\end{array}
	\right.
	\nonumber
\end{eqnarray}
In the limit of $a_*\to +\infty$, $r_{\rm ISCO}/M\to \sqrt{3}a_*$. 
In Fig. \ref{fig:superKerr_radius}, we show the event horizon $r_{\rm H}$ ({\it dotted line}) 
and the innermost stable circular orbit $r_{\rm ISCO}$ ({\it solid line}) as a function of the 
spin parameter $a_*$. For $a_*=8\sqrt{6}/3~(\sim 6.532)$, the radius of the ISCO becomes 
$6M$ which is the same value as that for the non-rotating black hole (i.e. $a_*=0$). The 
minimum value of $r_{\rm ISCO}=(2/3)M~(\sim 0.667 M)$ which is achieved at 
$a_*=a_{\rm cr}\equiv 4\sqrt{2}/(3\sqrt{3})~(\sim 1.089)$. 

\begin{figure}
\begin{center}
\vspace{+0mm}
\hspace*{-0mm}
\includegraphics[width=70mm]{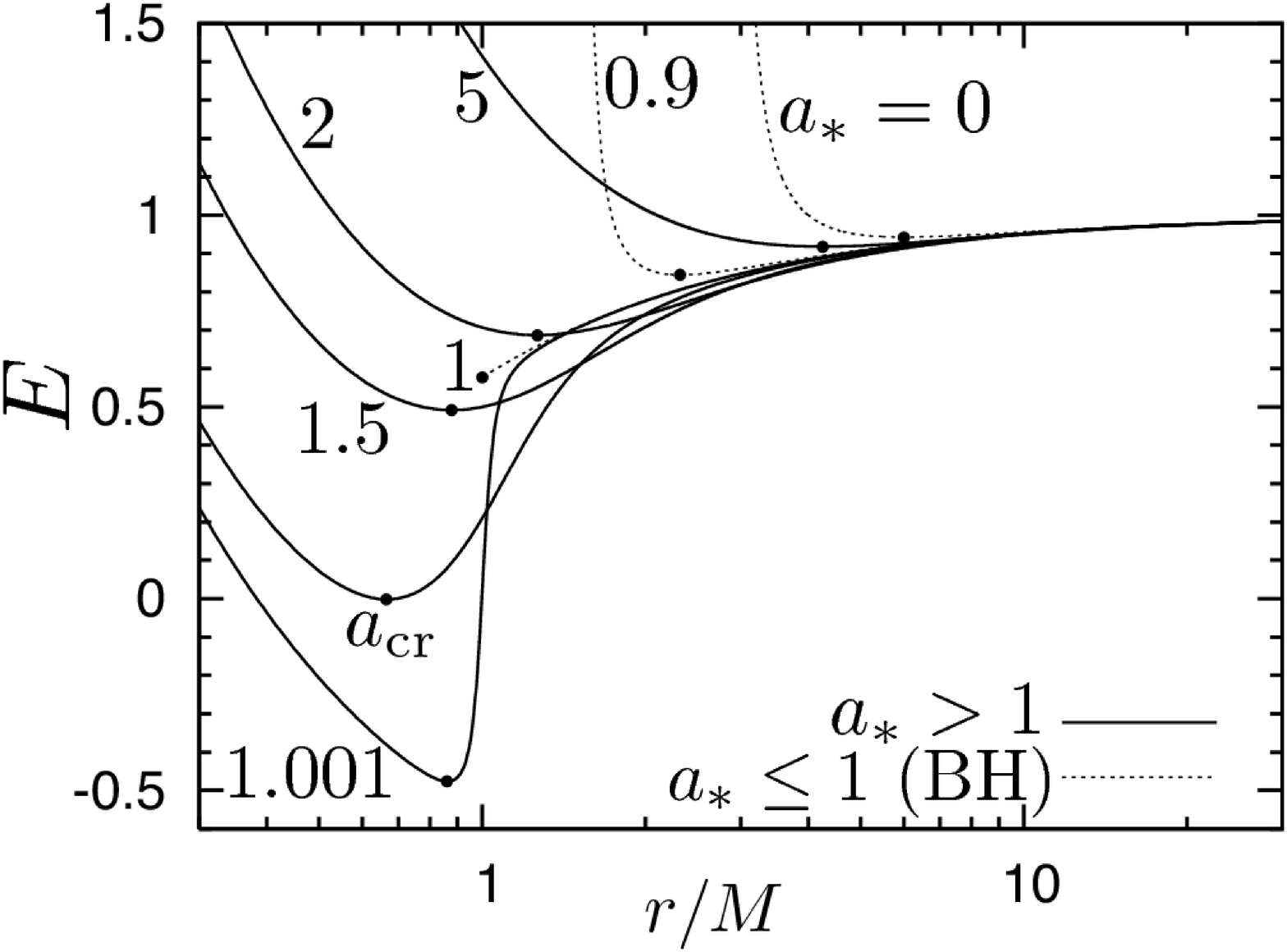}
\hspace{+5mm}
\includegraphics[width=71mm]{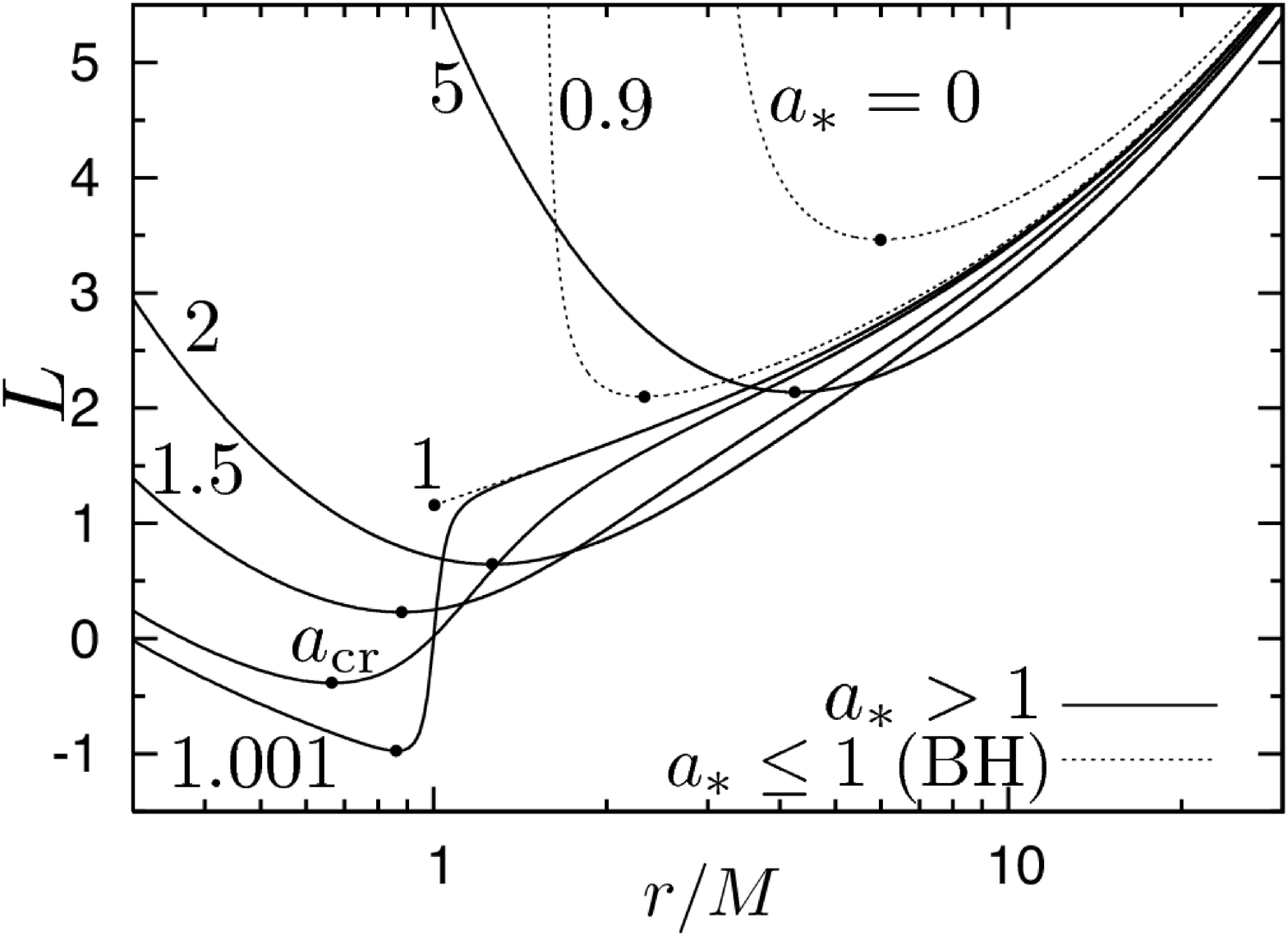}
\vspace{+0mm}\\
\hspace*{-5mm}
\includegraphics[width=72mm]{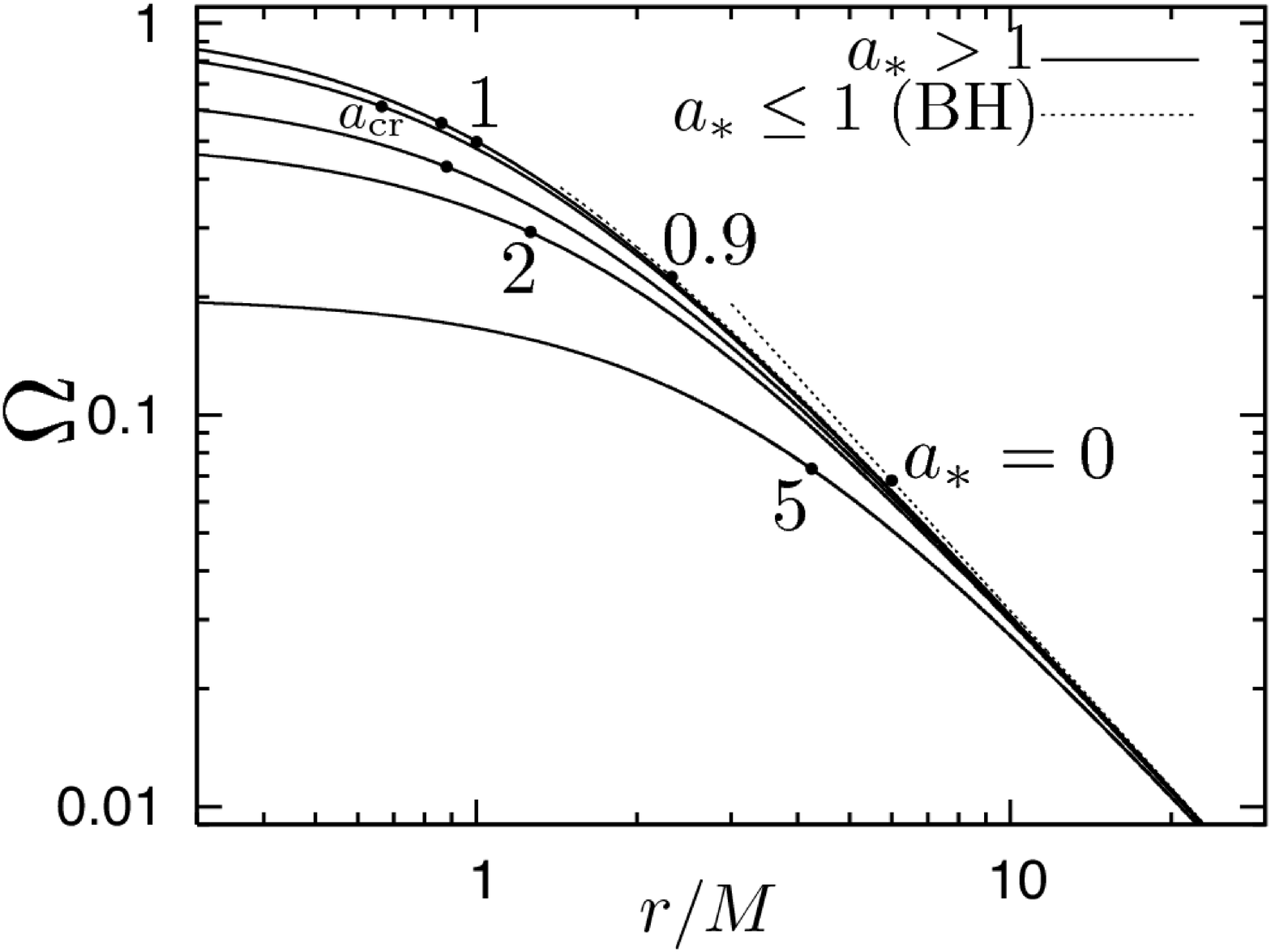}
\hspace{+0mm}
\includegraphics[width=72mm]{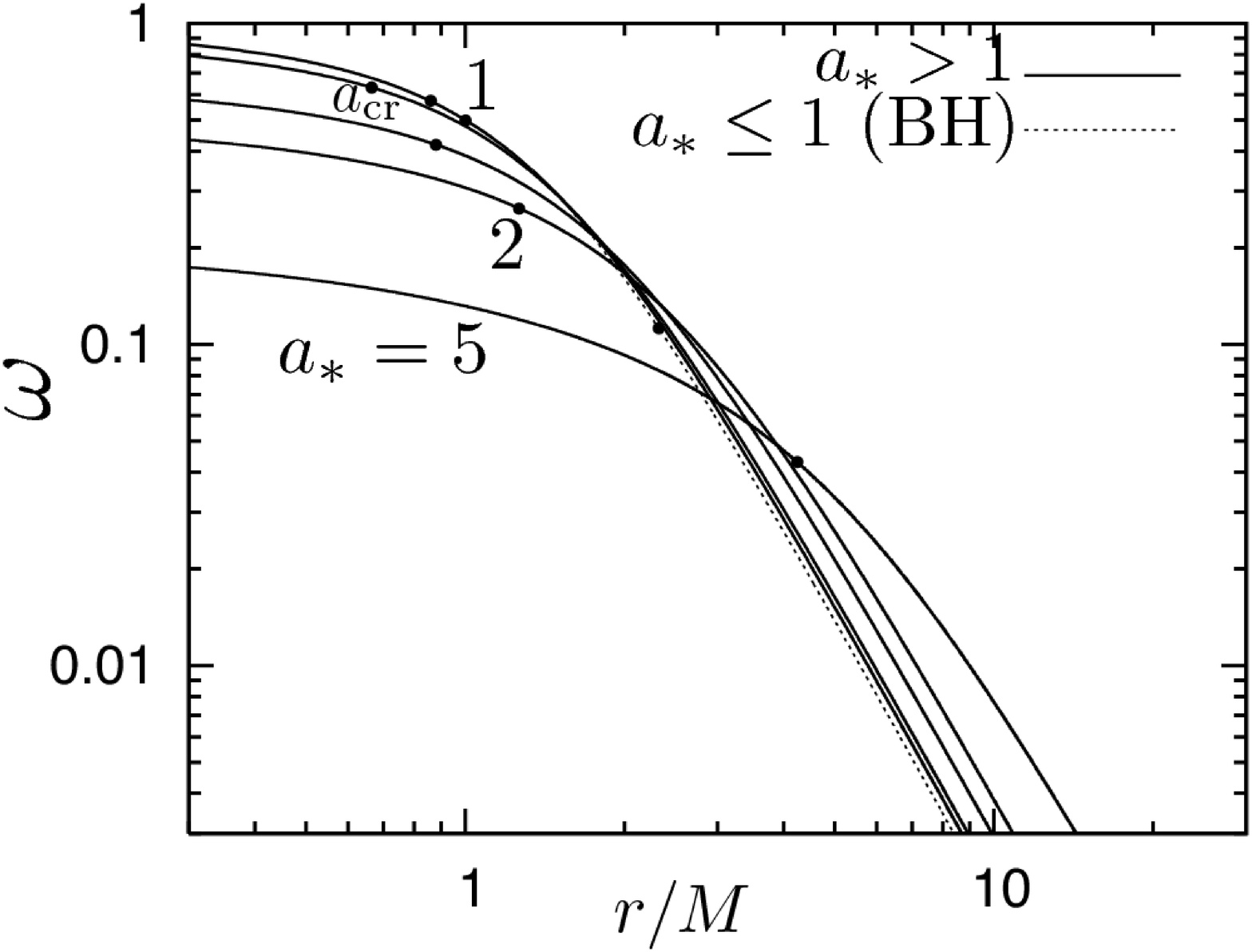}
\caption{\label{fig:superKerr_ELOm}
The total energy $E~(=-u_t)$ ({\it top left}), the angular momentum $L~(=u_\phi)$ 
({\it top right}), the angular velocity $\Omega~(=u^\phi/u^t)$ ({\it bottom left}) and the 
angular velocity of the frame-dragging $\omega$ ({\it bottom right}) of the orbiting 
particles around a super-spinar ({\it solid lines}) and 
a black hole (BH) ({\it dotted lines}). 
The values of $a_*$ are selected as 0, 0.9 and 1 for black holes and 1.001,  
$a_{\rm cr}=4\sqrt{2}/(3\sqrt{3})~(\sim 1.089)$,1.5, 2 and 5 for super-spinars. 
The radius of ISCO is denoted by the filled circles. 
}
\end{center}
\end{figure}

In Fig \ref{fig:superKerr_ELOm}, we plot the total energy $E~(=-p_t)$
({\it top left}), the angular momentum $L~(=p_\phi)$ ({\it top right}),
the angular velocity $\Omega~(=u^\phi/u^t)$ ({\it bottom left}) and the
angular velocity of the frame-dragging $\omega$ ({\it bottom right}) of
the orbiting particles around a super-spinar ({\it solid
lines}) and a black hole (BH) ({\it dotted lines}). Here, we have assumed
$s_1=s_2=1$ (positive energy at a large radius, and the prograde orbit). The values 
of $a_*$ are selected as 0, 0.9 and 1 for black holes and 1.001, 
$a_{\rm cr}=4\sqrt{2}/(3\sqrt{3})~(\sim 1.089)$, 1.5, 2 and 5 for super-spinars. 
The radius of ISCO is denoted by the filled circles. For the spin in the range of 
$1<a_*<a_{\rm cr}$, the energy at the ISCO becomes negative.

\begin{figure}
\begin{center}
\vspace{+0mm}
\hspace*{-0mm}
\includegraphics[width=140mm]{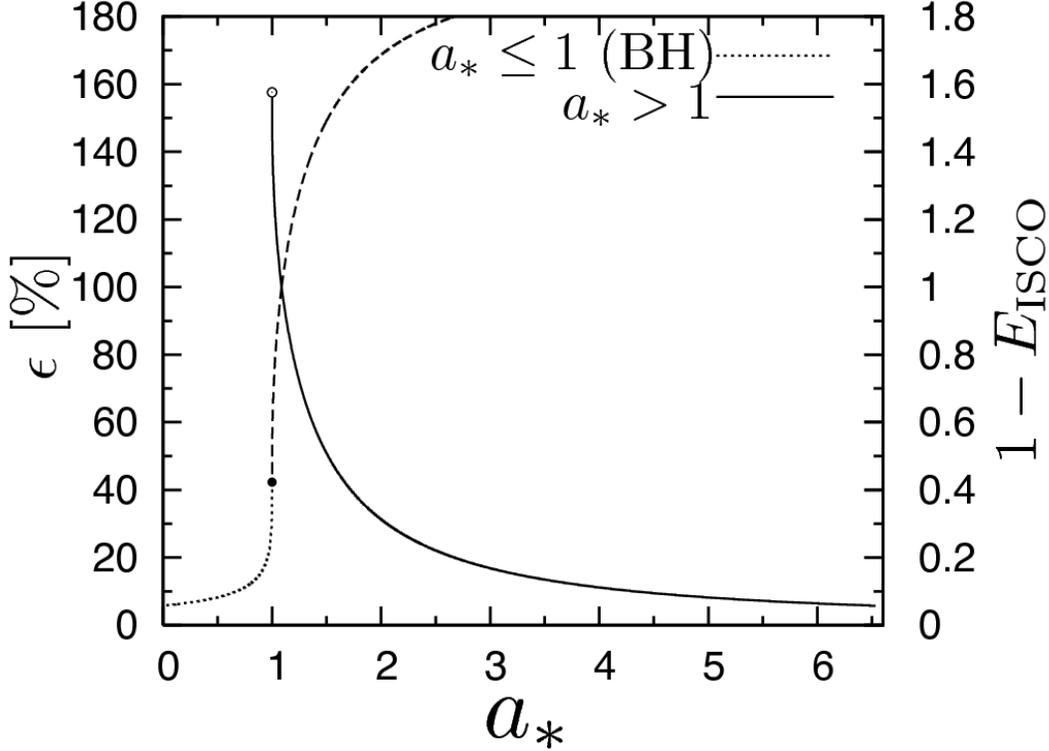}
\caption{\label{fig:efficiency}
The radiation efficiency $\epsilon~(=1-E_{\rm ISCO})$ with which the gravitational 
binding energy converts to the radiation energy when all the photons are escaping 
from the disc for the super-spinars with negative ({\it solid line}) and positive ({\it dashed line}) 
energy at ISCO and a black hole ({\it dotted line}). The locations of the maximally 
rotating black hole ({\it filled circle}) and the limit of maximum radiation efficiency 
for the super-spinar with the negative energy at ISCO ({\it blank circle}) are also shown. 
}
\end{center}
\end{figure}

The total radiation energy of the matter with the circular orbit in the equatorial plane 
is equal to the gravitational binding energy of the matter when it is at the ISCO. The 
efficiency $\epsilon$ with which the rest mass energy converts to the radiation energy 
of photons escaping from the disc is defined as the ratio of the rate of the radiation 
energy (=the rate of the gravitational binding energy) and the transportation rate of 
mass energy onto the central object. When all the emitted photons 
escape from the disc, 
for the matter at the ISCO, this efficiency is calculated by using the total energy $E$ 
at the ISCO as $\epsilon=1-E$. As is well known, for the non-rotating and the 
maximally rotating black holes these efficiencies become about 6\% and 42\%, 
respectively \cite{mtw73,fn98}.  In Fig. \ref{fig:efficiency}, we plot this 
efficiency, $\epsilon~(=1-E_{\rm ISCO})$, for the super-spinars with negative ({\it solid line}) 
and positive ({\it dashed line}) energy at ISCO and a black hole ({\it dotted line}) 
for the case that all the emitted photons escape from the disc. The locations of the 
maximally rotating black hole ({\it filled circle}) and the limit of maximum radiation 
efficiency for the super-spinar with the negative energy at ISCO ({\it blank circle}) are also 
shown in Fig \ref{fig:efficiency}. For the super-spinar (i.e., $1<a_*$), the efficiency decreases 
as $a_*$ increases. For the super-spinar with the spin within $1<a_*<a_{\rm cr}$, 
the radiation efficiency is over 100\%  (i.e., $1-E_{\rm ISCO}>1$ ). This is 
because the energy at ISCO is negative and then we can interpret that positive 
energy is extracted from the super-spinar \cite{df78, cn79, rt79, s81b}. It should be noted 
that the solution denoted by the dashed line in Fig \ref{fig:efficiency} for a super-spinar is 
not considered in this study. This is because this solution have negative energy at 
infinity and this is not the case for the accretion disc considered here. So, for a super-spinar, 
only the solution denoted by the solid line in Fig \ref{fig:efficiency} is considered in 
this study. The upper limit of $1-E_{\rm ISCO}$ is 
$(1-E_{\rm ISCO})_{\rm max}=1+1/\sqrt{3}~(\sim 1.577)$, i.e., the upper limit of the 
radiation efficiency is about 157.7\% as shown by the blank circle in Fig. 
\ref{fig:efficiency}. This is achieved for the super-spinar with the spin which is just 
above 1. The efficiency becomes 100\% at the spin $a_*=a_{\rm cr}$, where the 
minimum value of the radius of ISCO is achieved as $r_{\rm ISCO}=(2/3)M$. 
At the spin of $a_*=5/3~(\sim 1.667)$, the efficiency becomes $\sim 42\%$ which 
is the same value as for the maximally rotating black hole. At the spin of 
$a_*=8\sqrt{6}/3~(\sim 6.532)$, the efficiency becomes the same values as for the 
non-rotating black hole, i.e. 6\%. For the super-spinar with $a_*>8\sqrt{6}/3$, the radiation 
efficiency becomes smaller than that of the non-rotating black hole. It is noted that 
in these calculations we have assumed that all photons escape from the disc which 
is not completely realistic, i.e. in reality some photons emitted from the disc should 
be absorbed by the central object, not escaping into infinity \cite{t74}. 
Actually, in the vicinity of the black hole and the super-spinar, a large part of 
photons are trapped by the strong gravitational field of the central object. 
The analysis about this problem is presented in the past studies \cite{b73,s78,s81a}.  
However, it can be expected that the efficiency for the super-spinar with the spin of 
$1<a_*< 1.667$ is significantly larger than that of the 
maximally rotating black hole.

\section{Radiation Flux and Energy Spectrum of an Accretion disc}
\label{sec:spec}
The stress-energy tensor $T^{\mu\nu}$ is given by 
$T^{\mu\nu}=\rho_0 h u^\mu u^\nu+u^\mu q^\nu+u^\nu q^\mu+t^{\mu\nu}$, 
where $\rho_0$, $h$, $q^\mu$ and $t^{\mu\nu}$ are the rest-mass density, 
the relativistic enthalpy, the heat-flux tensor and the viscous tensor, respectively. 
For $q^\mu$ and $t^{\mu\nu}$, we have the orthogonality relations as 
$u_\mu q^\mu=0$ and $u_\mu t^{\mu\nu}=0$. For the stationary disc, the radiation 
flux $\mathcal{F}$ at the disc surface is calculated as $\mathcal{F}=q^\theta$. 
From the rest-mass conservation $\nabla_\mu (\rho_0 u^\mu)=0$, where 
$\nabla_\mu$ is the covariant derivative. The mass accretion rate $\dot{M}_0$ 
of the disc is calculated as $\dot{M}_0=-2\pi r \Sigma_0 u^r (={\rm constant})$, 
where $\Sigma_0$ is the surface density of the disc which is obtained by the 
integration of the rest-mass density along the disc thickness $2H$ as 
$\Sigma_0=\int^{H_\theta}_{-H_\theta}\rho_0~rd\theta$, where $H_\theta$ is 
the angular thickness of the disc, i.e. $H=r H_\theta$. Here, the radial 
component of the four-velocity is set to be negative, i.e. $u^r<0$.  From 
the energy conservation $\nabla_\mu T^{\mu\nu}h_\nu^{~t}=0$ and the 
angular momentum conservation $\nabla_\mu T^{\mu\nu}h_\nu^{~\phi}=0$, 
where $h^{\mu\nu}=g^\mu\nu+u\mu u^\nu$ is the projection tensor, we can 
obtain $\partial_r [\dot{M}_0 E + 2\pi r W^r_{~t}]=4\pi r \mathcal{F} E$ and 
$\partial_r [\dot{M}_0 L - 2\pi r W^r_{~\phi}]=4\pi r \mathcal{F} L$ where 
$W^{\mu\nu}$ is defined by the integration of the viscous tensor along the 
disc thickness as $W^{\mu\nu}=\int^{H_\theta}_{-H_\theta} t^{\mu\nu}~rd\theta$. 
From the orthogonality condition $u_\mu t^{\mu\nu}=0$, we have the relation 
$W^r_{~t}=-\Omega W^r_{~\phi}$. 

\begin{figure}
\begin{center}
\vspace{+0mm}
\hspace*{-0mm}
\includegraphics[width=140mm]{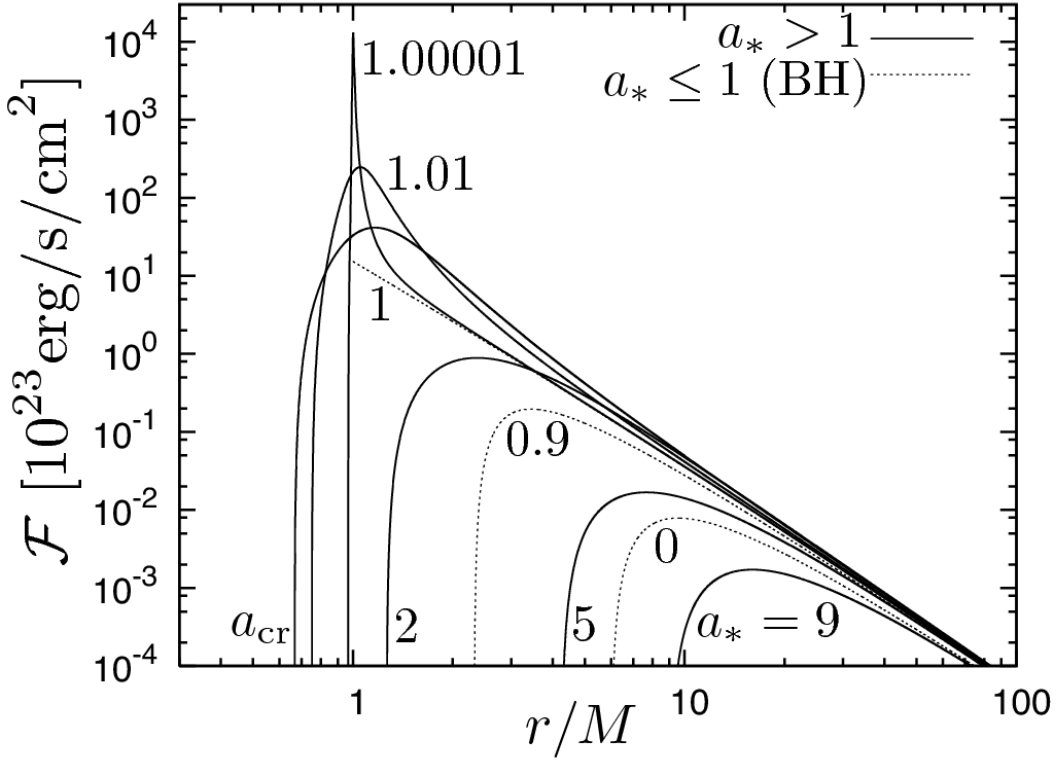}
\caption{\label{fig:PTflux}
The flux $\mathcal{F}$ [erg/s/cm$^2$] emitted from the geometrically thin and 
optically thick disc with the mass accretion rate $\dot{M}=0.1 \dot{M}_{\rm Edd}$ 
around a super-spinar ({\it solid lines}) and a black hole ({\it dotted lines}) 
with a mass $M=10M_\odot$. The values of $a_*$ are selected as 0, 0.9 and 1 
for black holes and 1.00001, 1.01, $a_{\rm cr}=4\sqrt{2}/(3\sqrt{3})~(\sim 1.089)$, 
2, 5 and 9 for super-spinars. 
}
\end{center}
\end{figure}

By using these relations,
the local flux at the disc $\mathcal{F}(r)$ [erg cm$^{-2}$ s$^{-1}$] is calculated 
as \cite{pt74}
\begin{equation}
\mathcal{F}(r)=\frac{\dot{M}_0}{4\pi r}f(r), \label{eq:flux}
\end{equation}
where $f(r)$ is given by 
\begin{equation}
f(r)=\frac{-\partial_r \Omega}{(E-\Omega L)^2}\int^r_{r_{\rm ms}} (E-\Omega L)~\partial_r L~dr. 
\end{equation}
Here, the inner boundary of the disc is set to be the radius of the ISCO 
inside which there is no torque and no radiation. For a Kerr black hole with 
$a_*<1$, the analytic expression for $f(r)$ is given by the Eq. (15n) in \cite{pt74}, 
which cannot be used for $a_*=1$. For a black hole with $a_*=1$, we can calculate $f(r)$ as 
\begin{equation}
f(r)=\frac{3}{2M}\frac{1}{x^2(x+2)(x-1)^2}\left[x-1-\frac{3}{2}\ln x+\frac{3}{2}\ln\left(\frac{x+2}{3}\right)\right]~~~({\rm for}~~~x>1),~~~
\label{eq:fa1}
\end{equation}
where $x=(r/M)^{1/2}$,
and for $x=1$, $f(r)=1/(3M)$. For a super-spinar, i.e. $a_*>1$, we can obtain the analytic form of $f(r)$ as 
\begin{eqnarray}
f(r)&=&\frac{3}{2M}\frac{1}{x^2(x^3-3x+2a_*)}
	\left\{
		x-x_0
		+\frac{3a_*^2}{x_*(x_*^2-3)}
		\ln(x/x_0)
	\right. \nonumber \\
	&&\left.
		-\frac{(x_*-a_*)^2}{x_*(x_*^2-1)}
		\ln\left(\frac{x-x_*}{x_0-x_*}\right) 
	\right. \nonumber \\
	&&\left.
		+\frac{1}{2(x_*^2-1)}
		\left(
			x_*-2a_*-\frac{2a_*x_*}{x_*^2-3}
		\right)
		\ln\left(\frac{x^2+x_*x+x_*^2-3}{x_0^2+x_*x_0+x_*^2-3}\right)
	\right. \nonumber \\
	&&\left.
		-\frac{6}{(x_*^2-1)\sqrt{3(x_*^2-4)}}
		\left(
			x_*^2/2-1+a_*x_1+\frac{a_*^2}{x_*^3-3}
		\right)
	\right.\nonumber\\
	&&\left.
		\left[
		\tan^{-1}\left(\frac{2x+x_*}{\sqrt{3(x_*^2-4)}}\right)-
		\tan^{-1}\left(\frac{2x_0+x_*}{\sqrt{3(x_*^2-4)}}\right)
		\right]
	\right\}
\label{eq:f}
\end{eqnarray} 
where 
\begin{eqnarray}
x_0&=&(r_{\rm ISCO}/M)^{1/2},\nonumber\\
x_*&=&-(a_*-\sqrt{a_*^2-1})^{1/3}-(a_*+\sqrt{a_*^2-1})^{1/3}.
\end{eqnarray}
Based on Eqs. (\ref{eq:flux}), (\ref{eq:fa1}) and (\ref{eq:f}), we can analytically 
calculate the flux $\mathcal{F}$ [erg/s/cm$^2$] emitted from the disc. In 
Fig \ref{fig:PTflux}, we plot the flux emitted from the geometrically thin and 
optically thick disc with the mass accretion rate $\dot{M}=0.1 \dot{M}_{\rm Edd}$ around 
a super-spinar ({\it solid lines}) and a black hole ({\it dotted lines}) with a mass $M=10M_\odot$. The 
values of $a_*$ are selected as 0, 0.9 and 1 for black holes and 1.00001, 1.01, 
$a_{\rm cr}=4\sqrt{2}/(3\sqrt{3})~(\sim 1.089)$, 2, 5 and 9 for the super-spinars. 
For all the calculations presented in Fig. \ref{fig:PTflux}, the radiation flux becomes 
zero within the radius of ISCO within which there is no torque. Then, the peak flux is 
achieved at a radius which is slightly larger than the radius of ISCO. 

As denoted in the previous section, the radiation efficiency for the super-spinar with the spin in the range of $1<a_*<5/3$ is larger than the radiation efficiency of the maximally rotating black hole (see Fig \ref{fig:efficiency}). In Fig \ref{fig:PTflux}, we can see that in the cases when a super-spinar with its spin in the range of $1<a_*<5/3$,  for any radius of the disc around such a super-spinar, the radiation flux is larger than that for the maximally rotating black hole. Then, for such a super-spinar, the total flux integrated along the disc surface also becomes larger than that of the maximally rotating black hole. For a super-spinar with the spin in the range $a_*>8\sqrt{6}/3$, the radiation efficiency becomes smaller than that of the non-rotating black hole as seen in the previous section. In the case of $a_*=9$ in Fig. \ref{fig:PTflux}, for any radius of the disc, the radiation flux is
  smaller than that for the non-rotating black hole ($a_*=0$). It is interesting that the amount of the local radiation flux for the super-spinar with the spin in $1<a_*<5/3$ can be larger than that of the maximally rotating black hole ($a_*= 1$) by a few orders of magnitude.  

\subsection{Temperature of the Accretion disc}
\label{sec:temp}

\begin{figure}
\begin{center}
\vspace{+0mm}
\hspace*{-0mm}
\includegraphics[width=140mm]{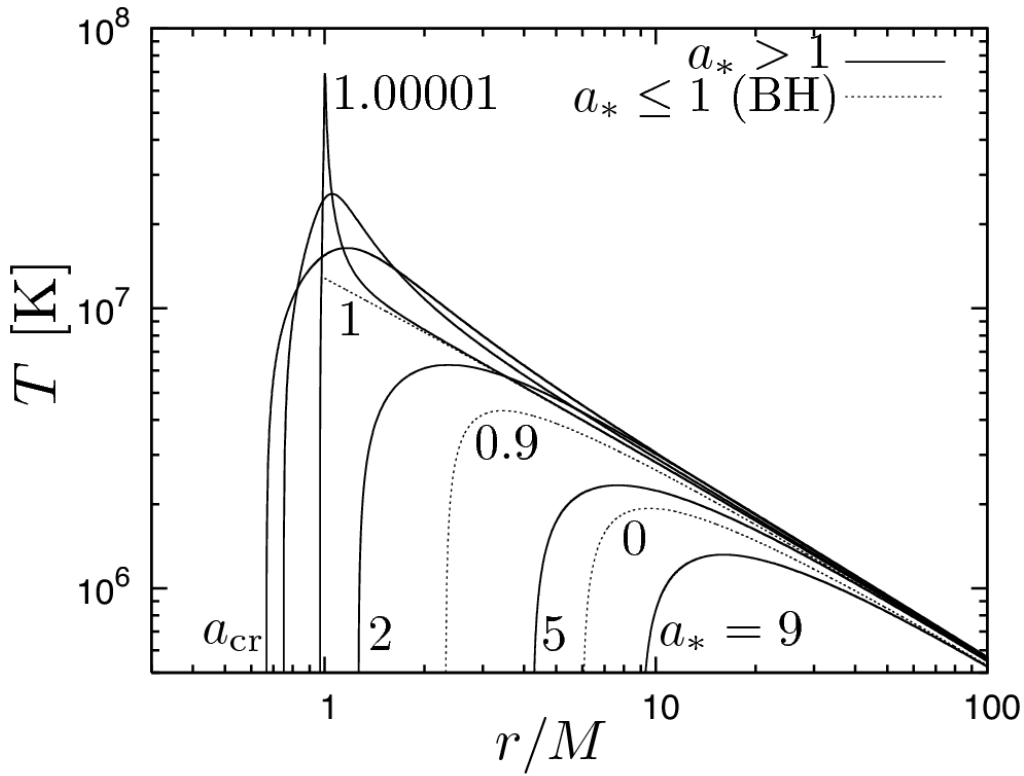}
\caption{\label{fig:T}
The radial profile of the effective temperature $T$ [K] [$=(F/\sigma)^{1/4}$] of the geometrically thin and optically thick disc for the same parameters used in Fig. \ref{fig:PTflux}. 
}
\end{center}
\end{figure}

By assuming that the local radiation spectrum of the disc surface obeys the blackbody spectrum, it is possible to calculate the effective temperature of the accreting matter. In this case, the effective temperature $T$ of the disc is related to the blackbody flux $\mathcal{F}$ with the Stefan-Boltzmann equation, $\mathcal{F}=\sigma T^4$, where $\sigma$ is the Stefan-Boltzmann constant ($\sigma=5.670\times10^{-5}$ erg s$^{-1}$ cm$^2$ K$^{-4}$). By using the local flux $\mathcal{F}$ calculated from Eqs. (\ref{eq:flux}), (\ref{eq:fa1})  and (\ref{eq:f}) and the Stefan-Boltzmann relation, the effective temperature of the disc can be calculated. In Fig \ref{fig:T}, we plot the radial profile of the effective temperature $T$ [K] [$=(F/\sigma)^{1/4}$] of the geometrically thin and optically thick disc around super-spinars ({\it solid lines}) and black holes ({\it dotted lines}) with the same parameters used in Fig. \ref{fig:PTflux}. 

In a similar manner as radiation flux of the disc in Fig \ref{fig:PTflux}, we can see in Fig \ref{fig:T} that in the cases when a super-spinar with its spin in the range of $1<a_*<5/3$,  for any radius of the disc around such a super-spinar, the temperature of the accreting matter is larger than that for the maximally rotating black hole. Also, for a super-spinar with the spin in the range $a_*>8\sqrt{6}/3$, the temperature of the accreting matters becomes smaller than that of the non-rotating black hole as seen in the case of $a_*=9$ in Fig. \ref{fig:T}.  It is noted that the temperature of the accreting matters for the super-spinar with the spin in $1<a_*<5/3$ becomes larger than that of the maximally rotating black hole ($a_*\sim 1$) by at most a several factor. 
The radiation efficiency of the particle at the radius of ISCO for $a_*=1.00001$ is larger than 
that for $a_*=1.01$ by about a factor 2 (see Fig. \ref{fig:efficiency}). In addition, the disc 
for $a_*=1.00001$ achieves much higher temperature than for the case of $a_*=1.01$ as shown 
in Fig. \ref{fig:T}. 

\subsection{Observed Energy Spectrum} 

\begin{figure}
\begin{center}
\vspace{+0mm}
\hspace*{-0mm}\includegraphics[width=140mm]{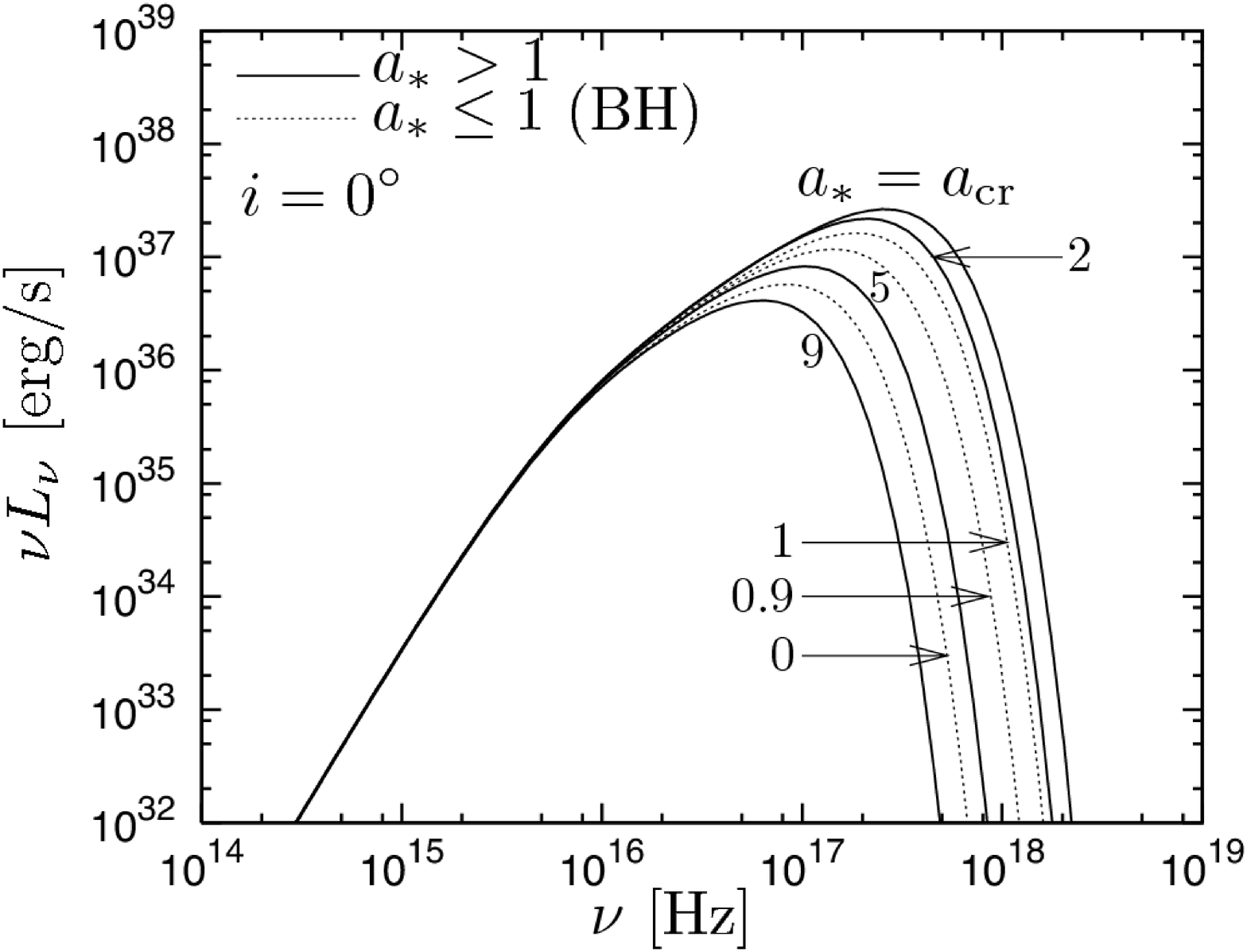}\vspace{3mm}\\
\hspace*{-0mm}\includegraphics[width=140mm]{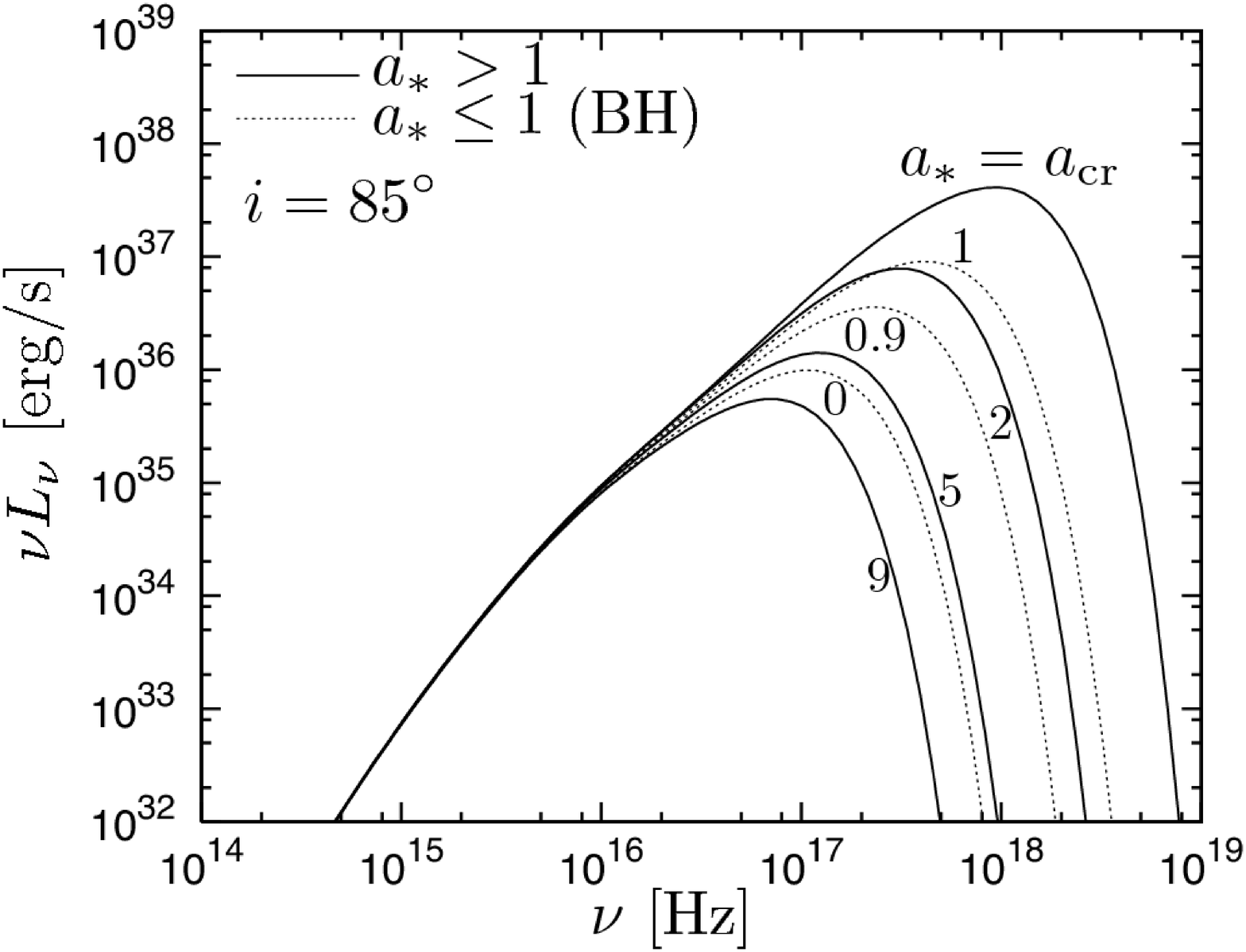}
\caption{\label{fig:SED}
The observed spectral energy distribution $\nu L_\nu$ [erg/s] of the geometrically thin and  
optically thick disc for the same parameters used in Fig. \ref{fig:PTflux} for the viewing angles 
$i=0^\circ$ and $85^\circ$.  
}
\end{center}
\end{figure}

Here, we calculate the energy spectrum of the disc observed by the distant observer which is the 
function of the mass of the central object $M$, the mass accretion rate $\dot{M}$, the spin of the 
central object $a_*$ and the viewing angle between the direction of the observer and the rotation 
axis of the disc $i$. The calculations are performed by solving general relativistic radiative 
transfer in the Kerr spacetime by including the usual special and general relativistic effects such 
as the Doppler boosting, the gravitational redshifts, the light bending and the frame-dragging. 
The calculation method of the observed energy spectrum $L_\nu$ [erg/s] is given in 
\ref{app:cm}, and this method is basically same as that used in \cite{tw07}. 
In Fig \ref{fig:SED}, we plot the spectral energy distribution $\nu L_\nu$ [erg/s] of the geometrically 
thin and optically thick disc  around super-spinars ({\it solid lines}) and black holes ({\it dotted lines}) for the 
viewing angles $i=0^\circ$ ({\it top}) and $85^\circ$ ({\it bottom}). The other parameters are same 
as those used in Fig \ref{fig:PTflux}. Here, we assume the outer radius of the disc as $r_{\rm 
max}=10^4 M$. 

As expected from the calculations given in Figs \ref{fig:PTflux} and \ref{fig:T}, the disc with a larger 
maximum temperature produces the energy spectrum extending to higher energy. The part of the 
lowest photon energy in the energy spectrum corresponds to the outer region (lowest temperature 
region) of the disc. So, for all the cases in Fig \ref{fig:SED}, we have the same energy spectrum in 
the part of the lowest photon energy \cite{kfm08}. For the case of $i=85^\circ$ in Fig \ref{fig:SED}, 
in the part of the middle photon energy in the energy spectrum, for a super-spinar with the spin around 
$1.01 \lesssim a_* \lesssim a_{\rm cr} $ the slope of the energy spectrum becomes slightly steeper 
than that of others. The slope in the middle part reflects the radial profile of the temperature (see, 
Fig \ref{fig:T}). In Fig \ref{fig:T} we can see that for super-spinars with spins of $a_*=1.01$ and $a_{\rm cr}$ 
have more increasing temperature profiles than other cases. These signatures can be seen in the 
middle part of the energy spectrum in the bottom panel of Fig \ref{fig:SED}. As a result, for a super-spinar 
with the spin of $1.01\lesssim a* \lesssim a_{\rm cr}$ and the large viewing angle, the slope of the 
middle part of the energy spectrum gives the characteristic signature of such a super-spinar. On the other 
hand, for other super-spinars (i.e. $1<a_*\lesssim 1.01$ and $a_{\rm cr}\lesssim a_*$) this signature 
cannot be seen clearly. 

In the same way as the radiation flux and the temperature of the disc in Fig \ref{fig:SED}, we can 
see that in the cases when a super-spinar with its spin in the range of $1<a_*<5/3$, for any energy of the 
photon emitted from the disc around such a super-spinar, the emitted energy becomes larger than that for 
a maximally rotating black hole. For a super-spinar with the spin in the range $a_*>8\sqrt{6}/3$, for any 
energy band the emitted energy becomes smaller than that of the non-rotating black hole. 
The disc for $a_*=1.00001$ produces photons with 
much higher energy. This is because the disc for $a_*=1.00001$ achieves much higher 
temperature than for the case of $a_*=1.01$.  

\subsection{Contributions of Energy Extracted from a Central Onject  in Energy Spectrum}

\begin{figure}
\begin{center}
\vspace{+0mm}
\hspace*{-0mm}
\includegraphics[width=140mm]{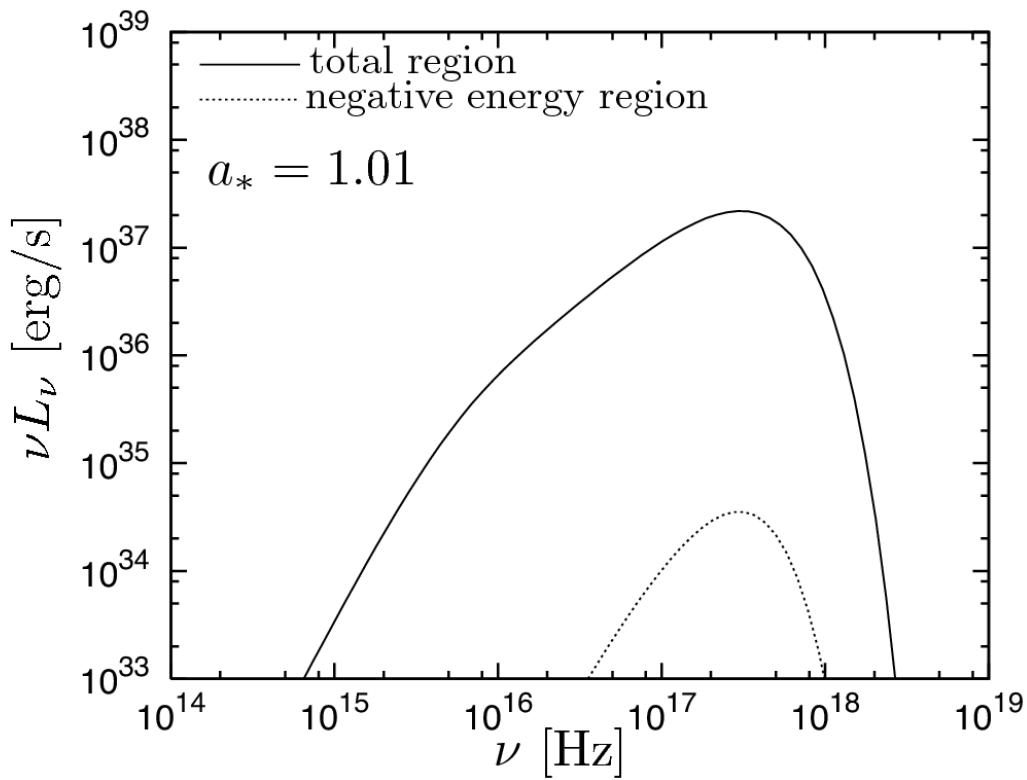}
\caption{\label{fig:SEDtpn}
The energy spectrum $\nu L_\nu$ [erg/s] for a super-spinar 
with the spin of $a_*=1.01$; the energy spectrum contributed 
from the whole region of the disc ({\it solid line}) and the energy spectra 
contributed from the positive energy region ({\it dashed line}) and 
the negative energy region ({\it dotted line}). The total energy 
spectrum is the sum of the energy spectra from the positive and the 
negative energy regions. 
}
\end{center}
\end{figure}

As already stated, the negative energy region appears in the accretion
disc around a super-spinar with the spin in the range of
$1<a_*<a_{\rm cr}=4\sqrt{2}/(3\sqrt{3})~(\sim 1.089)$. We next investigate the effects of
photons originated from the negative energy region in the energy
spectrum and the total radiation energy. In Fig \ref{fig:SEDtpn}, we
give the energy spectrum $\nu L_\nu$ [erg/s] for a super-spinar 
with the spin of $a_*=1.01$; the total energy spectrum
({\it solid line}), the energy spectra contributed from the positive
energy region ({\it dashed line}) and the negative energy region ({\it
dotted line}). The total energy spectrum is the sum of the energy
spectra of positive and negative energy regions. In Fig
\ref{fig:SEDtpn}, we can see that the contributions to the energy
spectrum from the emission originated from the negative energy region is
a relatively minor component. The total bolometric luminosity
$L_B^{\rm total}$ is calculated by the integration of the energy flux as
$L_B^{\rm total}=\int L_\nu~d\nu$. Separately, we can also calculate the
luminosity $L_B^-$ by the integration of the energy spectrum $L^-_\nu$
contributed by the negative energy region as $L^-_B=\int L^-_\nu~d\nu$,
where $L_\nu^-$ is the luminosity calculated from the photons emitted in
the negative energy region as shown by the dotted line in Fig
\ref{fig:SEDtpn}. 

\subsection{Can we confirm the Kerr bound from the thermal X-ray spectrum?}

\begin{figure}
\begin{center}
\vspace{+0mm}
\hspace*{-0mm}
\includegraphics[width=140mm]{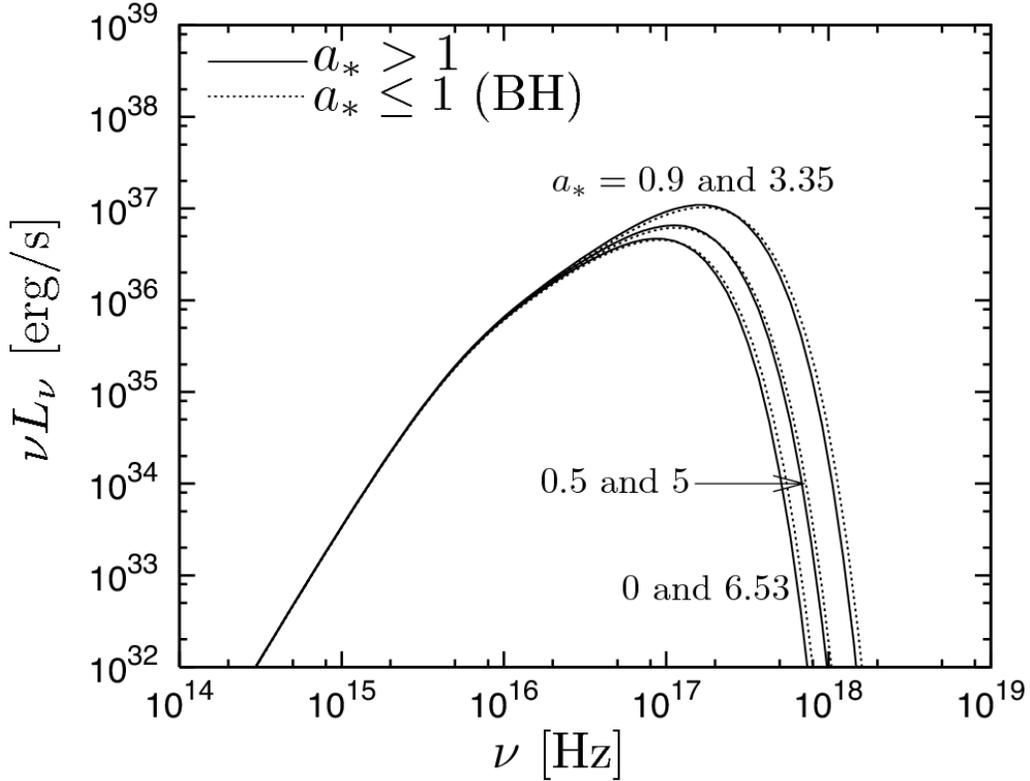}
\caption{\label{fig:similarSED}
The similar pairs of the energy spectra $\nu L_\nu$ [erg/s] of 
the geometrically thin and optically thick disc 
for a black hole ({\it dotted lines}) and a super-spinar ({\it solid lines}). 
The pairs of the spins for a black hole $a_*^{\rm BH}$ and its 
counterpart super-spinar $a_*^{\rm super-spinar}$ are 
$(a_*^{\rm BH},~a_*^{\rm super-spinar})=(0,~6.56)$, (0.5, 4.8) and (0.9,~3) ({\it left to right}).
}
\end{center}
\end{figure}

We finally give the examples of the very similar pairs of the energy
spectra of a black hole and a super-spinar. In Fig
\ref{fig:similarSED}, we plot the similar energy spectrum $\nu F_\nu$
[erg/s] of the geometrically thin and optically thick disc for 
a black hole ({\it dotted lines})
and 
a super-spinar ({\it solid lines}). 
The pairs of the spins for a black hole $a_*^{\rm BH}$ 
and a super-spinar $a_*^{\rm super-spinar}$ are $(a_*^{\rm BH},~a_*^{\rm super-spinar})=(0,~6.53)$, (0.5, 5) and (0.9,~3.35) 
({\it left to right}). As shown in this figure, surprisingly, for given black holes with some value of the spin $a_*(\le 1)$, we can always find its counterpart objects with the spin $a_*$ larger than the unity whose observed spectrum is very similar to and practically indistinguishable from that of the black hole. As a results, we can not confirm the Kerr bound only by using the X-ray thermal spectrum of the black hole candidates. 

\section{Discussion}
\label{sec:dis}
Here, we discuss about important topics relating to this study. 
After giving the discussion mainly in the research of the gravitational physics, 
we discuss about astrophysical topics. 

In this paper, we consider the object which is described by the Kerr metric but 
have specific angular momentum larger than its mass. For such object, 
no emission from the object is assumed and we implicitly assume the general 
relativity near curvature singularity is replaced by other gravitational theory 
described in the Introduction. These assumptions have the background as follows. 
Since the last century, extensive investigations have been performed on 
the research of black holes and naked singularities, around which the strong-field 
of gravity is achieved and the nonlinearity of gravitation is prominent. In particular, 
the cosmic censorship conjecture \cite{p69,p79,w97,p98} has been proposed but 
its proof is yet very limited. On the other hand, it has been revealed that there are 
many examples of solutions to the Einstein field equations which have naked 
singularities and contain physically reasonable matter fields,although no such 
example has been proven to be completely generic (see, e.g. \cite{harada04,joshi00}). 
This conjecture is often useful to deduce the properties of spacetimes and 
black holes (see, e.g. \cite{he73}). and hence many researchers on 
classical general relativity tend to assume it. However, from a quantum gravity 
point of view, the motivation is not so clear to believe that the cosmic 
censorship must hold within classical general relativity and, in fact, it has been 
pointed out that naked singularities in classical theory can be viewed as 
a window into new physics including quantum gravity~\cite{hn04}. 
If the specific angular momentum is greater than its mass in the Kerr solution, 
which is a stationary, axisymmetric and vacuum solution to the Einstein field 
equations, there is no horizon but naked singularity. However, it  might be 
reasonable to consider  that around the singularity classical general relativity 
is broken down and actually there is no singularity at the center \cite{bf09}. 
In such cases, some physical mechanism such as quantum gravity effects 
replace the singularity with some finite radius $R$. One can imagine 
the radius $R$ is very small but do not know how. Even within classical 
theory, the supercritically rotating Kerr solution might approximately describe 
the geometry exterior to a rapidly rotating compact object. In the past studies, 
based on quantum field theory in curved spacetime, the explosive emission 
from a forming naked singularity in gravitational collapse has been argued as 
a possible observational signature of naked singularities 
\cite{hwe82,bsvw98a, bsvw98b,hin00a,hin00b,hin02}. Although
these explosive signatures might appear in the observational signatures, 
in this study we assume no emission from the central objects.
We do not know whether these assumptions are reasonable or not, but the similar 
assumptions are used in the past studies (see the references in the following 
paragraphs).  

Before this study, there are a lot of past studies about the observational 
feasibility of the super-spinning objects and/or naked singularity in a 
variety of astrophysical contexts such as;  direct radio observations of 
a super-spinning Kerr object~\cite{bf09, hm09}, accretion disc around 
a super-spinning Kerr object~\cite{ts05}, gravitational lensing 
phenomena by a super-spinning Kerr object~\cite{ve02,vk08,gy08,wp07}, 
light rays from a forming naked singularity~\cite{d98,nki03}, 
particle creation (emission) from a forming naked singularity~
\cite{bsvw98a,bsvw98b,hin00a,hin00b,vw98,sv00}, 
gravitational radiation from a forming naked singularity~
\cite{nsn93,ihn99a,ihn99b,ihn00}, physical processes in naked 
singularity formation~\cite{hin02, jdm02,jgd04}, connection to 
gamma-ray bursts (GRBs)~\cite{jdm00,a99}. 

Most of the past calculations related to general relativistic
accretion disc models  assume a black hole as the central compact object
in the center of the accretion disc.  If the observational data contain
the physical information in the strong-field of gravity, these data can
be used to test the assumption of the black hole as the central object
and/or test the theory of gravitation including general relativity
in the strong-field regime.  Based on these motivations, recently several authors 
considered and/or calculated the accretion disc model around the central objects 
except the black hole in general relativity \cite{btb01,t02,ynr04,g06,pkh08,hkl08,phk08}. 
These calculations give the emissivity profiles of the discs surrounding the central objects 
including quark, boson, or fermion stars, wormholes and brane-world
black holes, or discs in $f(R)$ modified gravity models. 

It is widely accepted that most of the astrophysical black hole candidates discovered 
so far consist of the central object and the viscous accretion disc system. 
Therefore, in order to identify the objects with the super-spinning objects, 
it is essential to study the observational signatures of the viscous accretion 
disc around the assumed objects. Since the last century, 
a number of black hole candidates have been discovered by astronomical 
observations through the electromagnetic signatures (from radio to 
X-ray/gamma-ray) from the accreting matters plunging onto the central objects. 
Such radiation from the vicinity of the central objects contains the information 
about the space-time structure and the plasma in the strong gravitational field. 
Strictly speaking, to identify the central object with a black hole, we need to 
show not only that the observation is explained by the assumption that 
the central object is a black hole but also that it cannot be explained by 
the assumption that the central object is anything else. Although the test 
of gravitational theory is out of scope in the present paper, even general 
relativity is not a trivial assumption because it has never been so 
accurately tested in such a strong-field regime as around black holes. 
In this context, it is essentially important to clarify the relationships between 
the physics in the strong gravity and the observational features of 
the accreting plasma such as electromagnetic energy spectrum. 
In the context of the cosmic censorship, the observational identification 
of the central object provides a rather direct astrophysical test. 
For this purpose, it is at least required to find the distinguishable 
observational features between black holes and super-spinning objects 
(or naked singularities)~\cite{bf09,ve02,wp07}. 

As denoted in the Introduction, the X-ray spectrum of the black hole 
candidate generally consists of the thermal component originating 
from the accretion disc around the black hole and the non-thermal 
component originating from the high energy photons which are 
up-scattered in e.g. corona above or in the accretion disc. 
In this study, we only assumed the thermal component for simplicity. 
However, this is not valid especially for the hard X-ray 
spectrum. We know that there are many astrophysical objects with 
high energy radiation which can not be naturally explained by 
assuming the accretion disc used in this study. Especially, hard 
X-ray and gamma-ray radiation from the observed objects can 
not be simply explained by the standard  disc around a black hole. 
For the explanation of such observed high energy radiation, past 
researchers proposed many physical processes such as the 
inverse-Comptonization of the corona near the central object for 
the hard X-ray emission (e.g. \cite{st80,t94,ps96,mlm00,lms02, c09}),  
dark matter annihilation for X-ray and gamma-ray radiation 
(e.g. \cite{ubel02,cflm04,bfp06,hfd07,dhs08}), and other 
many processes \cite{an05}. Although these emissions are 
also expected around a super-spinning object, if exists, these are the topics for 
the future studies. In addition to the geometrically thin accretion 
discs considered in this paper, many kinds of the discs are 
proposed. Among the variety of types of the accretion discs/flows, 
most basic one is the so-called standard disc or Shakura-Sunyaev 
disc \cite{ss73}, whose state is achieved when the mass accretion 
rate $\dot{M}$ is sub-Eddington, i.e. $\dot{M}< L_{\rm Edd}/c^2$, 
where $L_{\rm Edd}$ is the Eddington luminosity given by 
$L_{\rm Edd}=1.25\times10^{39}(M/10M_\odot)$ erg/s. Here, 
$M$ is the mass of the central object. For  this mass accretion 
rate, the disc becomes  geometrically thin and optically thick. 
This standard disc can be applied to black hole candidates in 
black hole binaries and active galactic nuclei. The general 
relativistic version of the standard disc is given for the first time 
in \cite{nt73} and \cite{pt74}. The accretion discs assumed in this 
study belongs to this type. Based on the accretion disc theory, 
for the other range of the mass accretion rate, different forms of 
the disc structures are realized; for example, radiatively 
inefficient accretion flow or advection-dominated accretion 
flow for much smaller mass accretion rate, supercritical accretion 
disc or slim disc for super-critical mass accretion rate, 
hypercritical accretion disc or neutrino-dominated accretion 
flow  for hyper-critical mass accretion rate (for review, 
see e.g. \cite{kfm08}). For such accretion discs with a different mass 
accretion rate, since the physical processes in the disc and 
the equation of state are different, the resultant energy spectrum 
also becomes different from the results given in this paper. 
Especially, it is important to investigate the observational 
signatures of the radiatively inefficient accretion flow in the 
Galactic Center, where the direct imaging observations by 
the radio interferometers will be performed in the near future. 
It is expected that the direct imaging observations will 
determine the background spacetime geometry such as 
the spin parameter \cite{d08, b73,fma00,t04,t05,bl06,m07}. 
These studies will be performed in the future.   

\section{Conclusions}
\label{sec:con}
The observational confirmation of the Kerr bound directly suggests the existence of a black hole. 
In this study, in order to investigate testability of this bound by using 
the observed X-ray energy 
spectrum of black hole candidates, we first calculate the energy spectrum for the object whose 
spacetime geometry is described by the Kerr metric but 
whose specific angular momentum is larger than its mass, and then compare 
the results with that of a black hole. We call this object a super-spinar in this study. 
The optically thick and geometrically thin disc is assumed and only 
the thermal energy spectrum seen by the distant observer is calculated 
by general relativistic radiative transfer calculations including 
usual special and general relativistic effects such as Doppler boosting, 
gravitational redshift, light bending and frame-dragging. 
After calculating a disc structure such as velocity fields (Fig\ref{fig:superKerr_ELOm}) 
and radiation efficiency at ISCO (Fig\ref{fig:efficiency}), we have calculated 
energy flux radiated from the disc (Fig \ref{fig:PTflux}), disc temperature (Fig{\ref{fig:T}}) and 
observed energy spectrum (Fig \ref{fig:SED}). We use the new analytic formula 
for the radiation flux of a disc. As known in past studies, some energy is extracted from the central 
objects whose specific angular momentum is larger than its mass. We have investigated 
the influence of the extracted energy on the energy spectrum of a disc. Finally, we compare the 
energy spectra of a super-spinar and that of a black hole. 
In terms of the energy spectrum observed by a distant observer, we have obtained the 
following results:  
\begin{itemize}
\item For the super-spinar with $1<a_*<5/3\simeq 1.667$, 
higher energy photons are emitted from the disc than those from the disc 
around the maximally rotating black hole (see Fig\ref{fig:SED}). This 
signature can be seen especially for the cases with large viewing angles, 
e.g. the case with $i=85^\circ$ in Fig \ref{fig:SED}. 
\item For the super-spinar with $1.01\lesssim a_* \lesssim 1.1$ and its large viewing 
angle, the slope of the middle energy part of the energy spectrum becomes 
slightly steeper than that of the case of the black hole (see, e.g., the case 
for $i=85^\circ$ in Fig \ref{fig:SED}). 
\item The influence of the extracted energy from a super-spinar on energy spectrum is negligible 
(see, Fig \ref{fig:SEDtpn}). That is, most of the radiation energy comes from 
the accreting matters with positive energy even when the energy is maximally 
extracted from the super-spinar.  
\item For a given black hole, we can always find its super-spinning counterpart 
in the range $5/3<a_{*}<8\sqrt{6}/3$ whose observed spectrum is very similar 
to and practically indistinguishable from that of the black hole 
(see Fig \ref{fig:similarSED}). 
As a result, we conclude that to confirm the Kerr bound we need more than 
the X-ray thermal spectrum of the black hole candidates. 
Although in principle black holes and super-spinars can be distinguished 
by the detailed observations of the energy spectrum, 
the distinction between the black 
holes and the super-spinars only by the steady-state emergent spectrum 
becomes a severe challenge to the future observational facilities.  
\item For the super-spinar with $a_{*}>8\sqrt{6}/3 \simeq 6.532$, the total 
radiation energy of the disc is lower than the disc around the non-rotating 
black hole. 
\end{itemize}
As a result of this study, we found, surprisingly, that for a given black hole
we can always find its super-spinning counterpart whose observed 
spectrum is very similar to and practically indistinguishable from that of 
the black hole. Then, in order to confirm the Kerr bound 
we need more than the X-ray thermal spectrum of the black hole candidates. 



\section*{Acknowledgments}
The author would like to thank Akira Tomimatsu, Masaaki Takahashi, 
Cosimo Bambi, Naoki Isobe, Hitoshi Negoro, Makoto Miyoshi and Mareki Honma 
for valuable comments and discussion. The author also thank anonymous referees 
for useful comments and suggestions which improve the original manuscript. 
The authors are supported by the Grant-in-Aid for Scientific
Research Fund of the Ministry of Education, Culture, Sports, Science
and Technology, Japan [Young Scientists (B) 18740144 and 21740190 (TH);  
21740149 (RT)].

\appendix 
\section{Calculation method of the observed energy spectrum}
\label{app:cm}

In this appendix, we describe the calculation method of the observed energy spectrum in Kerr spacetime. 
After describing the calculation method of the null geodesics in Kerr spacetime, 
the formula for energy flux, image and energy spectrum observed by a distant 
observer are presented. It should be noted that the calculation method given here 
was essentially already presented in the past studies, e.g. \cite{cb73,c83,rb94}, for the black hole spacetime. 
We have developed these codes in the past studies, e.g. \cite{tw07, t04}. 
The same calculation methods can be applied to the case of the super-spinar.  

\subsection{Calculation of null geodesics in Kerr spacetime}
In the Kerr spacetime described by the Boyer-Lindquist coordinates given by Eq. (\ref{eq:metric}), 
the integral forms of null geodesics are calculated as \cite{mtw73,c68}
\begin{eqnarray}
s_\theta \int^\theta \frac{d\theta}{\sqrt{\Theta(\theta)}}=s_r \int^r \frac{dr}{\sqrt{R(r)}}, \label{eq:rth}\\
t = s_\theta \int^\theta \frac{-a(aE\sin^2\theta-L_z)}{\sqrt{\Theta(\theta)}}d\theta 
+ s_r \int^r \frac{(r^2+a^2)P}{\Delta\sqrt{R(r)}}dr,\\
\phi = s_\theta \int^\theta \frac{-(aE\sin^2\theta-L_z)}{\sin^2\theta\sqrt{\Theta(\theta)}} d\theta 
+ s_r \int^r \frac{aP}{\Delta\sqrt{R(r)}} dr
\end{eqnarray}
where $s_r=\pm 1$ and $s_\theta=\pm 1$ determine the direction of the integration along the geodesics and
$\Theta(\theta)$ and $R(r)$ are respectively given as 
\be
\Theta(\theta) &=& \mathcal{Q}-\cos^2\theta \left[a^2(m^2-E^2)+L_z^2/\sin^2\theta\right]\\
R(r)	&=& E^2r^4-(\mathcal{Q}+L_z^2-a^2E^2)r^2
		+2M[\mathcal{Q}+(L_z-aE)^2]r -a^2\mathcal{Q}. 
\ee
Here, $E(\equiv -p_t)$, $L_z(\equiv p_\phi)$ and $\mathcal{Q}$ are respectively energy, angular momentum 
with respect to the rotation axis and Carter constant, all of which are constants of motion. In this study, since we 
consider the axisymmetric disk in the equatorial plane, we only need to solve $r$ and $\theta$ components of 
the null geodesics given by Eq. (\ref{eq:rth}). If we introduce the variables, $u$ and $\mu$, defined as 
\be
u\equiv \frac{M}{r}~~~~~{\rm and}~~~~~\mu\equiv \cos\theta, 
\ee
and insert these variables into Eq. (\ref{eq:rth}), we obtain 
\be
s_\mu \int^\mu \frac{d\mu}{V_*(\mu^2)} = s_u \int^u \frac{du}{\sqrt{U_*(u)}}, \label{eq:muu}
\ee
where $s_\mu=\pm 1$ and $s_u=\pm 1$ determine the direction of the integration along the geodesics, and 
$V_*(\mu^2)$ and $U_*(u)$ are respectively defined as 
\be
V_*(\mu^2)&=&  -a_*^2\mu^4-[\eta+\zeta^2-a_*^2]\mu^2+\eta
	\\
U_*(u) &=& 
	-a_*^2\eta u^4+2[\eta+(\zeta-a_*)^2]u^3-(\eta+\zeta^2-a_*^2)u^2+1.
\ee
Here, $\eta$ and $\zeta$ are respectively defined as
\be
\eta \equiv \frac{\mathcal{Q}}{M^2E^2}~~~~~{\rm and}~~~~~\zeta \equiv \frac{L_z}{ME}. 
\ee
It is easily shown that we have the relation of 
\be
\eta+(a_*-\zeta)^2\ge 0
\ee
and the case of $\eta+(a_*-\zeta)^2=0$ requires $\theta$ to be constant (see, section 63 of \cite{c83}). 
Therefore, we do not require the case of $\eta+(a_*-\zeta)^2=0$. 
For all the possible values of $a_*$, $\eta$ and $\zeta$, the function $V_*(\mu^2)$ can have the forms of 
\be
V_*(\mu^2)=
a \mu^4+b \mu^2+c,~a\mu^2+b,~~{\rm or}~~a \label{eq:mutype}
\ee
where $a$, $b$ and $c$ are variables described by $a_*$, $\eta$ and $\zeta$. Especially, it is noted that 
$a$ is the non-zero variable. In a similar manner, the function $U_*(u)$ have the forms of 
\be
U_*(u)=a u^4+b u^3 +c u^2 +1~~{\rm or}~~a u^3 +b u^2 +1, \label{eq:utype}
\ee
where $a$, $b$ and $c$ are variables described by $a_*$, $\eta$ and $\zeta$. 
For each type of $V_*(\mu^2)$ and $U_*(u)$ given Eqs. (\ref{eq:mutype}) and (\ref{eq:utype}), 
we can solve the integrals of $\mu$ and $u$ by using the formula of the elliptic integrals, and the explicit 
forms of these solutions are given in \cite{rb94}. Our geodesic code is created based on these formula. 
On the other hand, the null geodesics can also be calculated by the numerical integrations of the 
Hamilton-Jacobi equations. We have also developed the null geodesic code for the arbitrary spacetime 
with 3+1 form, and we have confirmed that both codes using the analytic formula and the numerical integrations 
show the good agreement for the calculation of the null geodesics in Kerr spacetime.  

In the present study, we consider the distant observer. In such a case, two constants, $\eta$ and $\zeta$, used 
for the calculations of the null geodesics are related to the celestial coordinates $(x,~y)$ of the images seen by 
the distant observer as \cite{c83} 
\be
x = -\frac{\zeta}{\sin i}~~~{\rm and}~~~y=\pm \sqrt{\eta+a_*^2\cos^2i-\frac{\zeta^2}{\tan^2 i}} 
\ee
and inversely we have 
\be
\zeta = -x\sin i~~~{\rm and}~~~\eta=y^2+(x^2-a_*^2)\cos^2 i 
\ee
where $i$ is the viewing angle between the rotation axis of the black hole and the direction of the observer. 
By using these formula, we calculate $\eta$ and $\zeta$ from the celestial coordinates $(x,~y)$. 

\subsection{Calculation of the observed specific intensity}
Next, we describe the calculation method of the observed spectrum in Kerr spacetime which was established 
by the past studies, e.g. \cite{cb73}, for the black hole spacetime. For the super-spinar, the calculation method 
is essentially same.  In order to calculate the observed energy spectrum, first we have to calculate the observed 
specific intensity, $I_{\nu_{\rm obs}}$ [erg s$^{-1}$ cm$^{-2}$ str$^{-1}$ Hz$^{-1}$], of the accretion disk with the 
observed photon frequency $\nu_{\rm obs}$. The observed specific intensity is related to the specific intensity measured 
in the local rest frame of the disk as 
\be
I_{\nu_{\rm obs}} = g^3 I_{\nu_{\rm rest}}
\ee
where $I_{\nu_{\rm rest}}$ is the specific intensity in the local rest frame and $g$ is defined as 
\be
g \equiv \frac{\nu_{\rm obs}}{\nu_{\rm rest}} \label{eq:g}
\ee
where $\nu_{\rm rest}$ is the photon frequency measured in the local rest frame. 
In this study, since we consider the accretion disk of which the specific intensity in the local rest frame is 
given by the black body spectrum $B_{\nu_{\rm rest}} (T)$, where $T$ is the temperature of the accretion disk 
calculated in \S \ref{sec:temp}. The $g$-factor given in Eq. (\ref{eq:g}) is calculated by using 
the same method in \cite{cb73} as 
\be
g = \frac{(-p_\mu u^\mu)_{\rm obs}}{(-p_\mu u^\mu)_{\rm rest}} 
\ee  
where in the observer frame since we have 
\be
(u^t,~u^r,~u^\theta,~u^\phi)_{\rm obs} = (1,~0,~0,~0), 
\ee
then 
\be
(-p_\mu u^\mu)_{\rm obs} = -E, 
\ee
where $E$ is the constant of motion along the null geodesics.  On the other hand,  
in the local rest frame in the accretion disk, we calculate $p_\mu$ from the geodesic equations in Kerr spacetime 
\cite{c68} and $u^\mu$ is calculated from the Keplarian disk given in \S\ref{sec:disc}. In this manner, 
since we can calculate the $g$-factor which includes both the special and general relativistic effects such 
as the Doppler boosting, the gravitational redshifts, the frame-dragging effects, etc, the photon frequency 
in the local rest frame $\nu_{\rm rest}$ is calculated from $g$ and the observed photon frequency as 
$\nu_{\rm rest} = g^{-1}\nu_{\rm obs}$. This photon frequency measured in the local rest frame is a 
function of $a_*$, $i$, $x$, $y$ and $\nu_{\rm obs}$. 

\subsection{Calculation of observed flux, images and energy spectrum} 
By using the observed specific intensity given in the last subsection, the observed flux, $F_{\nu_{\rm obs}}^{\rm obs}$ 
[erg s$^{-1}$ cm$^{-2}$ Hz$^{-1}$], at the observed photon frequency $\nu_{\rm obs}$ is calculated as 
\be
F_{\nu_{\rm obs}}^{\rm obs} = \frac{1}{d^2} \int d\nu_{\rm rest}~g^4 I_{\nu_{\rm rest}} 
\delta(\nu_{\rm obs}-g\nu_{\rm rest}) 
\label{eq:flux}
\ee
where $\delta$ is the delta function and $d$ is the distance of the observer. The observed flux calculated in this 
manner is also the function of $a_*$, $i$, $x$, $y$ and $\nu_{\rm obs}$.  
Based on the observed flux calculated in Eq. (\ref{eq:flux}), we can calculate the observed images for given 
observed photon frequency $\nu_{\rm obs}$, the spin $a_*$ and the viewing angle $i$. 
In the observed images, in addition to the direct images of the accretion disk, we take into account the indirect images 
which is formed along the null geodesics penetrating the equatorial plane inside the ISCO. 

By calculating the observed images of the accretion disk for some appropriate range of the observed 
frequency and integrating the observed flux in $x$-$y$ plane, we can calculate the observed luminosity, 
$L_{\nu_{\rm obs}}$ [erg s$^{-1}$ Hz$^{-1}$], at the observed photon frequency $\nu_{\rm obs}$, i.e. 
\be
L_{\nu_{\rm obs}} = \int dx \int dy~F_{\nu_{\rm obs}}^{\rm obs},  
\ee 
where in terms of the integration ranges of $x$ and $y$ we consider geodesics penetrating the equatorial plane 
inside the maximum radius, $r_{\rm max}$, of the accretion disk. In the main part of this paper, we denote the 
observed photon frequency as $\nu$ for simplicity.


\section*{References}
\begin{thebibliography}{10}

\bibitem{mtw73}
Misner C. W., Thorne K. S., Wheeler J. A., 1973, {\it Gravitation} (W. H. Freeman and Company, New York, 1973)

\bibitem{fn98}
Frolov V. P., Novikov I. D., 1998, {\it Black Hole Physics: Basic Concepts and New Developments} (Kluwer Academic Publishers, 1998)

\bibitem{gh09}
Gimon E. G., Horava P., 2009, Phys. Lett. B672, 299

\bibitem{bf09}
Bambi C., Freese K., 2009, Phys. Rev. D., 79, 3002 

\bibitem{bft09}
Bambi C., Freese K., Takahashi R., 2009,  the proceedings of "21th Rencontres de Blois: Windows on the Universe" (Blois, France, 21-27 June 2009) (arXiv:astro-ph/0908.3238)

\bibitem{p69}
Penrose R., 1969, Riv. Nuovo Cim.  1, 252 

\bibitem{p79}
Penrose R., 1979, in {\it General Relativity and Einstein Centenary Survey}, edited by S. W. Hawking and W. Israel (Cambridge University Press, Cambridge, England, 1979)

\bibitem{w97}
Wald R. M., , 1997, "Gravitational Collapse and Cosmic Censorship", arXiv:gr-qc/9710068

\bibitem{p98}
Penrose R., 1998, in {\it Black Holes and Relativistic Stars}, edited by R. M. Wald (University of Chicago Press, Chicago, 1998), p.103

\bibitem{ss04} 
Sekiguchi Y., Shibata M., 2004, Phys. Rev. D, 70, 084005

\bibitem{st91}
Shapiro S., Teukolsky S., 1991, Phrs. Rev. Lett., 66, 994

\bibitem{st92}
Shapiro S., Teukolsky S., 1992, Phrs. Rev. D, 45, 2006 

\bibitem{harada04}
Harada T., 2004, Pramana, 63, 741 

\bibitem{t65} 
Thorne K. S., 1965, {\it Geometrodynamics of Cylindrical Systems}, Ph.D. Thesis, Princeton University

\bibitem{t72} 
Thorne K. S., 1972, Nonspherical gravitational collapse: A short review, in Magic Without Magic, ed. by J. R.. Klauder (Freeman , San. Francisco), 231-258 

\bibitem{b02} 
Berger B. K., 2002, Living Reviews in Relativity, 5, 1

\bibitem{cpcc08a} 
Cardoso V., Pani P., Cadoni, M., Cavaglia M., 2008, Classical and Quantum Gravity, 25, 195010

\bibitem{cpcc08b} 
Cardoso V., Pani P., Cadoni M., Cavaglia M., 2008, Phys. Rev. D, 77, 14044 

\bibitem{pccc09}
Pani P., Cardoso V., Cadoni M., Cavaglia M., 2009, the proceedings of Black Holes in General Relativity and String Theory - August 24-30 2008 - Veli Losinj, Croatia (ArXiv:gr-qc/0901.0850)

\bibitem{t74}
Thorne K. S., 1974, ApJ, 191, 507 

\bibitem{js09} 
Jacobson T., Sotiriou T. P., 2009, ArXiv:gr-qc/0907.4146

\bibitem{bfh09}
Bambi C., Freese K., Harada T., Takahashi R., Yoshida N., 2009, arXiv: gr-qc/0910.1634

\bibitem{s05} 
Shen Z.-Q., Lo K. Y., Liang M.-C., Ho P. T. P., Zhao J.-H., 2005, Nature, 438, 62

\bibitem{d08}
Doeleman S., et al., 2008, Nature,  455, 78 

\bibitem{b09} 
Broderick A. E., Fish V. L., Doeleman S. S., Loeb A., 2009, ApJ, 695, 59

\bibitem{y09} 
Yuan Y.-F., Cao X., Huang L., Shen Z.-Q., 2009, ApJ, 699, 722

\bibitem{h09} 
Huang L., Takahashi R., Shen Z.-Q., 2009, ApJ in press (ArXiv:astro-ph/0909.3687)

\bibitem{bl09} 
Broderick A. E., Loeb A., 2009, ApJ, 703, L104

\bibitem{hm09}
Hioki K., Maeda K., 2009, arXiv:astro-ph/0904.3575		       

\bibitem{zcc97}
Zhang S. N., Cui W., Chen W., 1997, ApJ, 482, L155

\bibitem{gmne01}
Gierli\'{n}ski M., Maciolek-Niedzwiecki A., Ebisawa K., 2001, MNRAS, 325, 1253 

\bibitem{dbht05}
Davis S. W., Blaes O. M., Hubeny I., Turner N. J., 2005, MNRAS, 621, 372 

\bibitem{msnrdl06}
McClintock J. E., Shafee R., Narayan R., Remillard R. A., Davis S. W., Li L.-X., 2006, ApJ, 652, 518

\bibitem{smndlr06}
Shafee R., McClintock J. E., Narayan R., Davis S. W., Li L.-X., Remillard R. A., 2006, ApJ, 636, L113 

\bibitem{rm06}
Remillard R. A., McClintock J. E., 2006, ARA\&A, 44, 49 

\bibitem{nms08}
Narayan R., McClintock J. E., Shafee R., 2008, "Astrophysics of Compact Objects"  eds. Y. F. Yuan, X. D. Li, D. Lai, AIP Conference Proceedings, 968, 265 (2008) (ArXiv:astro-ph/0710.4073)

\bibitem{mns08}
McClintock J. E., Narayan R., Shafee R., 2008, in {\it Black Holes} eds. M. Livio and A. Koekemoer (Cambridge University Press, 2008) (ArXiv:astro-ph/0707.4492)

\bibitem{df74}
de Felice F., 1974, Astron. Astrophys., 34, 15 

\bibitem{df78}
de Felice F., 1978, Nature, 273, 429 

\bibitem{cn79}
Calvani M., Nobili L., 1979, Nuovo Cimento, 51B, 247 

\bibitem{rt79}
Reina C., Treves A., 1979, ApJ, 227, 596 

\bibitem{s80}
Stuchl\'{i}k Z., 1980, Bull. Astron. Inst. Czechosl., 31, 129

\bibitem{s81a}
Stuchl\'{i}k Z., 1981, Bull. Astron. Inst. Czechosl., 32, 40

\bibitem{s81b}
Stuchl\'{i}k Z., 1981, Bull. Astron. Inst. Czechosl., 32, 68

\bibitem{bsb89}
Bi\v{c}\'{a}k J., Stuchl\'{i}k Z., Balek V., 1989, Bull. Astron. Inst. Czechosl., 40, 65

\bibitem{nt73}
Novikov I. D., Thorne K. S., 1973, in {\it Black Holes}, edited by C. DeWitt and B. DeWitt (Gordon and Breach, New York, 1973), p. 343

\bibitem{pt74}
Page D. N., Thorne K. S., 1974, ApJ, 191, 499

\bibitem{bpt72} 
Bardeen J. M., Press W. H., Teukolsky S. A., 1972, ApJ, 178, 347 

\bibitem{c68}
Carter B., 1968, Phys. Rev., 174, 1559

\bibitem{s08}
Shafee R., McKinney J. C., Narayan R., Tchekhovskoy A., Gammie C. F., McClintock J. E., 2008, ApJ, 687, L25

\bibitem{b73}
Bardeen J. M., 1973, in {\it Black Holes}, edited by C. DeWitt and B. DeWitt (Gordon and Breach, New York, 1973), p. 215

\bibitem{s78}
Stuchl\'{i}k Z., 1978, unpubl. Doctoral Thesis (The Charles University, Prague, in Czech)

\bibitem{tw07} 
Takahashi R., Watarai K., 2007, MNRAS, 374, 1515

\bibitem{kfm08}
Kato S., Fukue J., Mineshige S., 2008, {\it Black-Hole Accretion discs, Towards a New Paradigm} (Kyoto University Press, Kyoto, 2008)

\bibitem{joshi00}
Joshi P. S., 2000, Pramana, 55, 529 

\bibitem{he73}
Hawking S. W., Ellis G. F. R., 1973, {\it The Large Scale Structure of Space-Time}, (Cambridge University Press, Cambridge, 1973)

\bibitem{hn04} 
Harada T., Nakao K., 2004, Phys. Rev. D,  70, 041501(R)

\bibitem{hwe82}
Hiscock W. A., Williams L. G., Eardley D. M., 1982, Phys. Rev. D,  26, 751 

\bibitem{bsvw98a}
Barve S., Singh T. P., Vaz C., Witten L., 1998, Nucl. Phys. B, 532 361

\bibitem{bsvw98b}
Barve S., Singh T. P., Vaz C., Witten L.,1998, Phys. Rev. D, 58, 104018 

\bibitem{hin00a}
Harada T., Iguchi H., Nakao K., 2000a, Phys. Rev. D, 61, 101502R 

\bibitem{hin00b}
Harada T., Iguchi H., Nakao K., 2000b, Phys. Rev. D, 62, 084037

\bibitem{hin02}
Harada T., Iguchi H, Nakao K., 2002, Prog. Theor. Phys., 107, 449

\bibitem{ts05}
Torok G., Stuchlik Z., 2005, A\& A 437, 775 

\bibitem{ve02}
Virbhadra K. S., Ellis G. F. R., 2002, Phys. Rev. D,   65, 103004 

\bibitem{vk08}
Virbhadra K. S., Keeton C. R., 2008, Phys. Rev. D,  77, 124041 

\bibitem{gy08}
Gyulchev G. N., Yazadjiev S. S., 2008, Phys. Rev. D,  78, 083004 

\bibitem{wp07}
Werner M. C., Petters A. O., 2007, Phys. Rev. D,  76, 064024

\bibitem{d98}
Dwivedi I. H., 1998, Phys. Rev. D,  58, 064004 

\bibitem{nki03}
Nakao K., Kobayashi N., Ishiharam H., 2003, Phys. Rev. D,  67, 084002

\bibitem{vw98}
Vaz C., Witten L., 1998, Phys. Lett. B, 442, 90 

\bibitem{sv00}
Singh T. P., Vaz C., 2000, Phys. Rev. D,  61, 124005 

\bibitem{nsn93}
Nakamura T., Shibata M., Nakao K., 1993, Prog. Theo. Phys., 89, 821

\bibitem{ihn99a}
Iguchi H., Harada T., Nakao K., 1999a, Prog. Theo. Phys., 101, 1235 

\bibitem{ihn99b}
Iguchi H., Harada T., Nakao K., 1999b, Prog. Theo. Phys., 101, 1252

\bibitem{ihn00}
Iguchi H., Harada T., Nakao K., 2000, Prod. Theo. Phys., 103, 53 

\bibitem{jdm02}
Joshi P. S., Dadhich N., Maartens R., 2002, Phys. Rev. D,  65, 101501R 

\bibitem{jgd04}
Joshi P. S., Goswami R., Dadhich N., 2004, Phys. Rev. D,  70, 087502 

\bibitem{jdm00}
Joshi P. S., Dadhich N. K., Maartens R., 2000, Mod. Phys. Lett. A, 15, 991

\bibitem{a99}
Antia H. M., 1999, Gen. Rel. Grav., 31, 1675 

\bibitem{btb01}
Bhattacharyya S., Thampan A. V., Bombaci I., 2001, A\& A, 372, 925 

\bibitem{t02}
Torres D., Nucl. Phys., 2002, B626, 377

\bibitem{ynr04}
Yuan Y. F., Narayan R., Rees M. J., 2004, ApJ, 606, 1112 

\bibitem{g06}
Guzman F. S., 2006, Phys. Rev. D,  73, 021501

\bibitem{pkh08}
Pun C. S. J., Kov\'{a}cs Z., Harko T., 2008, Phys. Rev. D,  78, 024043

\bibitem{hkl08}
Harko T., Kov\'{a}cs Z., Lobo F. S. N., 2008, Phys. Rev. D,  78, 084005 

\bibitem{phk08}
Pun C. S. J., Harko T., Kov\'{a}cs Z., 2008, Phys. Rev. D,  78, 084015 

\bibitem{st80}
Sunyaev R. A., Titarchuk L. G., 1980, A\& A, 143, 374 

\bibitem{t94}
Titarchuk L., 1994, ApJ, 434, 570 

\bibitem{ps96}
Poutanen J., Svensson R., 1996, ApJ, 470, 249

\bibitem{mlm00}
Meyer F., Liu B. F., Meyer-Hofmeister E., 2000, A\& A, 361, 175 

\bibitem{lms02}
Liu B. F., Mineshige S., Shibata K., 2002, ApJ, 572, L173

\bibitem{c09}
Cao X., 2009, MNRAS, 394, 207 

\bibitem{ubel02}
Ullio P., Bergstrom L., Edsjo J., Lacey C. G., 2002, Phys. Rev. D,  66, 123502 

\bibitem{cflm04}
Cesarini A., Fucito F., Lionetto A., Morselli A.. 2004, Ullio, Astropart. Phys., 21, 267 

\bibitem{bfp06}
Bergstrom L., Fairbairn M., Pieri L., 2006, Phys. Rev. D,  74, 123515

\bibitem{hfd07}
Hooper D., Finkbeiner D. P., Dobler G., 2007, Phys. Rev. D,  76, 083012

\bibitem{dhs08}
Dodelson S., Hooper D., Serpico P. D., 2008, Phys. Rev. D,  77, 063512

\bibitem{an05}
Aharonian F., Neronov A., 2005, ApJ, 619, 306 

\bibitem{ss73}
Shakura N. I. , Sunyaev R. A., 1973, A\& A, 24, 337 

\bibitem{b73}
Bardeen J. M., 1973, in {\it Black holes}, edited by C. DeWitt and B. DeWitt (Gordon and Breach, New York, 1973), p. 215 

\bibitem{fma00}
Falcke H., Melia F., Agol E., 2000, ApJ, 528, L13 

\bibitem{t04}
Takahashi R., 2004, ApJ, 611, 996 

\bibitem{t05}
Takahashi R., 2005, PASJ, 57, 273

\bibitem{bl06}
Broderick A. E., Loeb A., 2006, ApJ, 636, L109 

\bibitem{m07}
Miyoshi M., et al. 2007, 2007, Publ. Natl. Astron. Obs. Japan,10, 15

\bibitem{cb73}
Cunningham C. T., Bardeen J. M., 1973, ApJ, 183, 237 

\bibitem{c83}
Chandrasekhar S., 1983, The Mathematical Theory of Black Holes (Clarendon Press, Oxford and Oxford University Press, New York) 

\bibitem{rb94}
Rauch K. P., Blandford R. D., 1994, ApJ, 421, 46

\end{thebibliography}
\end{document}